\documentclass[preprint,showpacs,preprintnumbers,amsmath,amssymb]{revtex4-2} 
\usepackage{graphicx}
\usepackage{bm}
\usepackage{color}
\usepackage{xcolor}
\usepackage{ragged2e}
\usepackage{subcaption}
\usepackage{amsmath}
\usepackage{amssymb}   
\usepackage{epsfig}
\usepackage{float}
\usepackage{needspace}
\usepackage{comment}
\usepackage{booktabs}   
\usepackage{multirow}   
\usepackage{float}      
\usepackage[colorlinks=true, allcolors=blue]{hyperref}

\DeclareGraphicsExtensions{.png,.pdf}
\usepackage[english]{babel}   
\usepackage{microtype}       
\usepackage{float}

\usepackage{array}
\newcolumntype{M}[1]{>{\centering\arraybackslash}m{#1}}

\newcommand{\red}[1]{\textcolor{red}{#1}}

\begin{document}   

\title{On the synaptic matrix eigenvalues  of sparsely connected neural networks}
\author{Mohd. Gayas Ansari}
\author{Pragya Shukla}
\affiliation{ Department of Physics, Indian Institute of Technology, Kharagpur-721302, West Bengal, India \\\\
E-Mail: shukla@phy.iitkgp.ac.in}
\date{\today}     

\widetext

\begin{abstract}

The spectral behaviour of the  synaptic matrix,  representing the neuronal connection strengths, is an important tool to  analyze the stability and transient dynamics  of a typical brain as well as its learning process and memory capacity. The complexity of the brain due  to large number of neurons as well as underlying transient mechanisms e.g. homeostasis, seizure or synaptic plasticity can lead to networks with time-varying degree and type of sparsity. This renders an exact determination of the synaptic matrix not only technically difficult but also meaningless, leaving its statistical analysis as the best available theoretical approach.    This motivates us to pursue a spectral analysis  of the synaptic matrix models  with different type of sparsity and thereby analyze latter's role on various aspects of network dynamics and stability. Our results have potential relevance for detemining the type of synaptic sparsity required to induce a specific brain function or desired transient mechanism e.g  for pharmacological effects or physiological modulators.

\end{abstract}
\maketitle
\section{Introduction}

Understanding  brain function  requires a detailed analysis  of the synaptic connectivity in neuron circuits as well as its relationship with the cognitive properties responsible for brain activity. A synapse is the  link that  brings presynaptic neuron cell membranes  into close proximity to the  postsynaptic ones and the neurons  forward signals to individual target cells. In a standard theoretical approach,   assuming  the dominant connectivity  to be pairwise, the circuit dynamics is associated with a matrix structure, namely, the synaptic matrix with  its entries representing synaptic strengths   between various pairs of neurons in the network.  Intense investigations of wide ranging neural networks, through  experimental, theoretical as well as numerical route have now well established the undeniable  relevance of the synaptic matrices in brain networks. A thorough understanding of these matrices, their structural dependence on the network of interest and  a detailed insight of their role in network dynamics is  therefore   highly desirable and is  primary objective of the present work.

A real brain contains large groups of neurons, with a single neuron potentially connected to many of them. The connections in a network can therefore be significantly high and may lead to emergence of a higher-level intelligence that cannot be observed from a single neuron. But the synaptic connectivity in a typical brain has restricted spatial range, is specific to neuronal type, location, and response properties and is essentially dynamic.  Indeed, synaptic plasticity, an  important neurochemical foundations of learning and memory, can cause the synapses to strengthen or weaken over time as their activity increases or decreases \cite{Citri2008}. Previous studies based on simple network models indicate that the plasticity mechanisms influences  synaptic variability  more significantly than the  mean synaptic strengths and can lead to networks with varying degree and type of sparsity. Indeed the complexity  of the  neural networks often renders exact determination of the synaptic connections not only quite challenging  but also meaningless due to their ever changeable nature. This leave a statistical modelling of the synaptic matrices, sensitive to network parameters, as the only viable option. The type of randomness associated with matrix elements e.g. Gaussian, log-normal, poisson is expected to be sensitive to the network. For example, a previous study of paired recordings in a layer 5 cortical neurons suggest a log-normal distribution for synaptic strengths \cite{buzsaki2014log}.  In addition, the synaptic matrices are often found to be sparse with many non-random entries or local structures (motifs) too \cite{Song2005}.

A synaptic matrix is in general a real asymmetric matrix, subject to additional constraints formulated on the basis of typical neuron dynamics observed in experiments. Based on the Dale's law, the individual neurons can either be excitatory (E) or inhibitory (I). This in turn implies existence of two type of columns (or rows) in a synaptic matrix, those with all positive elements (referred as excitatory columns) or  all negative elements (referred as inhibitory columns).   Biologically, experiments show that cortical neurons are constantly bombarded by
thousands of large excitatory and inhibitory signals, rising and
falling simultaneously, so their net effect is small and highly variable. This leaves the 
network in a high-input, near-cancellation regime, with large E and I inputs mostly offsetting each other,
leaving only small fluctuations.  In many cases e.g. awake cortex (mouse, cat, monkey), neurons fire irregularly and sparsely. The firing is  saturated If excitation were much larger than inhibition and is silenced if inhibition dominates \cite{vanvreeswijk1996}. The fact that firing is irregular with high variability supports the
idea that E and I are dynamically balanced, leaving only small net fluctuations to drive
the irregular, variable spiking patterns observed in the living brain. To prevent the excitatory feedback from producing runaway activity and maintain balance, presence of sufficient inhibition is also necessary \cite{vogels2005neural}. Indeed a precise balance of excitation and inhibition  has been demonstrated in spontaneous network \cite{haider2006neocortical}.

A state with excitation almost balancing inhibition is  crucial for the brain: it avoids both runaway excitation and silence, allows rapid responses to weak inputs, generates irregular spiking patterns, reduces
metabolic cost, and ensures survival of a wide range of independent activity modes. In contrast, without balance, a single outlier mode dominates the  spectrum, destabilizing dynamics and producing either overly regular synchronous firing or excessive suppression, both of which poorly match experimental data. While the precise information about the exact mechanism that lies behind this balancing act   still remains elusive, it is expected to manifest through constraints on the  synaptic matrix elements.  Indeed, as suggested by some previous studies, a near cancellation on an average of  excitatory and inhibitory inputs manifests through  the zero sum rules on the rows (columns) of the connectivity matrix (later referred as the {\it balanced condition} or {\it zero sum rule}). Based on previous studies, the role of row sum constraint is to prevent the emergence of strong outliers and retain the spectrum in a compact bulk around zero and thereby ensuring network's stability. But, as discussed later in the text, our analysis reveals an almost insensitivity of more generic synaptic matrix ensembles, mimicking sparsity in the network,  to the row sum constraint. The latter  therefore seems to be irrelevant in general for various stages of network dynamics where appearance of such ensembles is  expected more than the dense ones.

A randomization of the synaptic matrix manifest through randomaization of the activation variables e.g. firing rates, thus making it necessary to study their the statistical behavior. The latter can be derived from that of the eigenvalues of the synaptic matrix  which can appear along real line as well as on the complex spectral plane  as complex conjugate pairs. (Although under specific system conditions, known as  exceptional points, the matrix become non-diagonalizable but  such conditions are rare). The spectrum of the synaptic matrix provides a compact way to understand the collective behavior of large neural networks.  For example, the existence of spontaneous activity depends on whether the real parts of any of the eigenvalues are large enough to destabilize the silent state, the eigenvalues with large real parts also provide strong clues about the nature of the spontaneous activity in the full, nonlinear models. In addition, the eigenvalues provide deep geometric and statistical insight too: geometrically, large values correspond to directions in which activity is stretched while small values correspond to nearly collapsed directions.

In addition to the matrix constraints, the spectral statistics of the synaptic matrix depends also on the ensemble constraints  that represents it.  Indeed synaptic connections  are usually non-uniformly random but are spatially organized, with connection probability and strength decaying with distance. Pyramidal cells, the primary source of excitation, typically form dense, local connections, while many classes of inhibitory interneurons have broader axonal arbors, creating a center surround receptive field structure \cite{Barbour2007,buzsaki2014log,Douglas2004,Markram2004}. In addition a variation of the synaptic variability strength can lead to variation of the activated modes and thereby change of dynamics.  The external stimuli due to interaction with environment can also randomly affect the connections over time. In contrast, the switching on a controlled  extrenal stimuli  may cause the neurons to learn from the inputs and thereby  readjust their synaptic weights. It is then natural to query as to how the statistical properties of  synaptic matrices vary if and when the network parameters change due to internal or external causes.   This in turn requires analysis of a wide range of multiparametric ensembles of synaptic matrices, with network information e.g. sparsity, connectivity etc contained in ensemble parameters. In addition, the matrix constraints e.g. those due to Dale's law and balanced condition make these ensembles different from the standard real non-Hermitian matrix ensembles and, as a consequence, the available theoretical formulation for the statistical behavior of the latter can not be applied to former. While this renders a statistical formulation of the associated  synaptic matrix very desirable, the theoretical analysis is technically challenging.   
To circumvent the latter, previous studies considered the decomposition of the synaptic matrix into a simple, unstructured random matrix ($J$) and a separate mean-connectivity matrix ($M$). The balance condition forcing the row sums of the fluctuation matrix $J$ to zero is then a specific tool that makes the eigenvalues of the full matrix ($J+M$) identical to those of the simpler, analytically solvable $J$ matrix. The balance condition forcing the row sums of the fluctuation matrix J to zero is then a specific tool that makes the eigenvalues of the full matrix ($J+M$) identical to those of the simpler, analytically solvable $J$ matrix. The above tool however does not help much in case of sparse network.

The present work takes a more direct and biologically grounded approach  investigating network stability and dynamics. Instead of decomposing the matrix into abstract mathematical components, we consider a single, unified synaptic matrix where all relevant biological properties are integrated from the outset. In our model, Dale's Law, the mean synaptic strengths, and the variance of connections are all intrinsic properties of a single, physically viable matrix. This method allows for the creation of proper statistical ensembles of plausible network configurations, moving beyond the study of analytical abstractions that are, by design, distant from biological reality. We explore this unified construction in four distinct and progressively more realistic scenarios. 
Our first model is a synaptic matrix ensemble with a constant ratio of the excitatory off-diagonals to diagonals variance and similarly for inhibitory ones;  here the excitatory and inhibitory connections describe a complex network with long range uniform connectivity. Notwithstanding its simple matrix structure, the ensemble represents  a  synaptic matrix without relying on the J+M decomposition.

In general synaptic connections in the cortex are not uniformly random but are 
spatially organized, with connection probability and strength decaying with distance between basis states (later referred as "{\it spatial distance}). Pyramidal cells, the primary source of excitation, typically form dense, local connections, while many classes of inhibitory interneurons have broader axonal arbors, creating a center surround receptive field structure. In addition, the  stability in the brain is not achieved by a static, pre-calculated balancing act. Rather, it is the result of homeostatic plasticity and dynamic feedback, where inhibitory circuits constantly and actively track and stabilize excitatory activity. This ongoing process maintains the network in a high-conductance, "balanced" state, preventing runaway activity. To capture this  biological reality, it is necessary to define separate spatial scales for excitation ($b_e$) and inhibition ($b_i$); this motivates our second model i.e a synaptic matrix ensemble in which the variance of a synaptic connection is a function of the distance between the presynaptic and postsynaptic neurons, creating a synaptic matrix with a highly realistic, non-uniform statistical structure.

The two models mentioned above are based on  a "dense" network where every neuron could theoretically influence every other neuron, even though the connection strength may be relatively weak or varying with their separation. In general, however, the neural connectivity in real brains is extremely sparse to conserve energy and space. A single neuron connects to only a tiny fraction of the total population, strictly limited to its immediate physical neighbors. To capture this most fundamental constraint of biological wiring i.e sparsity, we  investigate another ensemble consisting of sparse synaptic matrices. The sparsity here is 
enforced by a strict interaction range ($R$), any connection beyond this range is set to exactly zero. This creates a highly sparse synaptic matrix, mimicking the efficient, low-cost wiring of real neural circuits like the hippocampus. By combining this strict sparsity with Dale’s Law, we can test whether network stability is driven by the vast majority of 'missing' connections (zeros) rather than the active ones.
Our findings in this spatially structured case are particularly revealing, as they demonstrate that the standard balance condition is insufficient to guarantee stability in the presence of strong spatial structure and E/I asymmetry, leading to the emergence of powerful outlier eigenvalues that the classical theory cannot predict. By directly modelling these more faithful representations of cortical wiring, this work demonstrates that achieving network stability is a more complex challenge than suggested by simpler, unstructured models, highlighting the limitations of established balancing techniques and opening the door for the development of more robust solutions.

We proceed as follows. With primary focus on the spontaneous activity of the neural network, we review the basic ideas  in section II and also introduce the generic form of the multivariate Gaussian ensemble $\rho(J)$ of the synaptic matrices $J$ which is used later to model the ensembles with different type of sparsity.  The section III gives the details of four specific cases of $\rho(J)$   as well as the numerical analysis of their spectral statistics. The section also discusses the implications of various features of the spectral distributions for network dynamics i.e the different ways the network can behave e.g. as a silent state, firing irregularly, or firing together in a rhythm, a seizure state. As these behaviors depend on the excitation/ inhibition strengths,  the input arrival rates and  the delay time a signal takes to travel from one neuron to another, they leave their fingerprints on the eigenvalue distribution through synaptic matrix elements. We conclude in section IV with a discussion of our main results.

\section{Theoretical Model}

The neuronal activity  described by standard models for a network of $N$ interconnected neurons  e.g. firing rate networks or spiking networks can be formulated in general by dynamical equations of type $\tau_i \, {{\rm d}x_i \over {\rm d}t} = F \bigg(x_1, \ldots, x_N \bigg)$ with  $x_i$  characterizing the activation variable of neuron $i$ and the function $F$ varying from one network to other. The stability of the network can then be analyzed by the fixed point analysis of the above set of equations i.e from the eigenvalues of the Jacobian matrix $J$  at the fixed point. The network at the fixed point is stable  if all eigenvalues have negative real parts thereby implying a small perturbation decaying  exponentially  in time.  The  instability of the network (i.e some perturbations growing exponentially) is indicated if some of the eigenvalues have a positive real part. Fixed points can be further
classified as stable or unstable nodes, unstable saddle points, stable or unstable spiral points, or stable or unstable improper nodes.

 To elucidate the connection between a Jacobean and a synaptic matrix, here we consider a simple firing rate network.  Assuming the input from a given presynaptic neuron  proportional to its firing rate, the total synaptic input can then be obtained by summing the contributions of all the presynaptic neurons (referred as firing rate network).  For a network of $N$ interconnected neurons, with neuron $i$ characterized by a firing rate  $x_i$, this leads to  \cite{vogels2005neural}
\begin{eqnarray}
\tau_i \, {{\rm d}x_i \over {\rm d}t} = - x_i + F \bigg(\sum_j J_{ij} \, x_j + I_i + \Theta(x_i) \bigg).
\label{eq1}
\end{eqnarray}
with $F$ as a  firing-rate function. Here $J_{ij}$ describes the connection from presynaptic neuron $j$ to postsynaptic neuron $i$ with excitatory connection correspond to $J_{ij} >0$, inhibitory connections to $J_{ij} < 0$ and $I_i$ refers to input term. The standard theta function $\Theta(x_i)$ acts as a current bias, inducing spontaneous firing or suppressing it: $\Theta(x_i) =1$ if $x_i >0$ else $0$. The time constant $\tau_i$ determines the variability of firing rate.  

Following from eq.(\ref{eq1}), the $N \times N$ synaptic matrix $J$ turns out to be the Jacobean matrix;
the stability analysis of the network can then be pursued in terms of the eigenvalues of $J$
(also referred as modes).

\subsection{Ensemble density}

Notwithstanding detailed statistical information available about the cortical network connections, the specific information about $J_{ij}$ entries is not available. Indeed different generic activity patterns of neurons  are expected to be best represented by  $J$-matrices with different distribution of  their entries. Based on standard statistical hypothesis e.g. maximum entropy hypothesis \cite{Shukla2002,Shukla2001,Ansari2024,Ansari2024complexity}, the lack of information can then be modelled by maximizing the distribution under known constraints.  This motivates us to consider following ensemble density
\begin{eqnarray}
\rho(J; v, b) = C \; \exp\left[-\sum_{k,l} \frac{\left(J_{kl}-b_{kl}\right)^2}{v_{kl}}\right] \;  f_1(J) \, f_2(J),
\label{rho}
\end{eqnarray}
with $v_{kl}$ and $b_{kl}$ as the arbitrary variance and mean value of the matrix element $J_{kl}$. As the limit $v_{kl} \to 0$ corresponding to a non-random $J_{kl}$ taking a value $b_{kl}$, the above density can model a wide range of random synaptic matrices including sparse as well as banded ones. 
Here $f_1$ imposes the positive definite condition on the $N_E$ excitatory columns and  $f_2$ the negative definite condition on the $N_I$ inhibitory columns (with $N=N_E + N_I$),
\begin{eqnarray}
f_1(J) &=& \bigg(\prod_{k=1}^{N_f} \prod_{l=1}^N \Theta(J_{kl}) .\bigg) \\
f_2(J) &=& \bigg(\prod_{l=N_f+1}^{N} \prod_{k=1}^N  (1-\Theta(J_{kl}) )\bigg).
\label{cons0}
\end{eqnarray}
The ensemble density in eq. (\ref{rho})  is later referred as the {\it unbalance conditioned}  or just {\it unbalanced} ensemble  for brevity.

The ensemble in eq.(\ref{rho}) describes a random synaptic matrix for a network without  
balancing of the excitation and inhibition inputs taken into account. 
But, as mentioned in previous section, the real life brain networks can display an almost balance 
of excitation and inhibition   \cite{Vogels2011}, both inputs very strong but
nearly cancelling each other on average (latter referred as the {\it balanced condition}). 
As mentioned in section I,  this appears as a {\it row sum zero constraint}) on the matrix elements 
in an arbitrary row and thereby affects the ensemble density, changing it to following form,
\begin{eqnarray}
\rho(J; v, b) = C \; \exp\left[-\sum_{k,l} \frac{\left(J_{kl}-b_{kl}\right)^2}{v_{kl}}\right] \;  f_1(J) \, f_2(J) \, f_3(J)
\label{rhob}
\end{eqnarray}
where
\begin{eqnarray}
f_3(J) = \prod_{k=1}^N \delta \left(\sum_{l=1}^N J_{kl}  \right) 
\label{cons1}
\end{eqnarray}
The ensemble density in eq.(\ref{rhob})  is later referred as the balance conditioned  Gaussian ensemble (BCGE).  A special case of eq.(\ref{rhob}) was analyzed in \cite{rajan2006eigenvalue}. In next section, we numerically analyze four different variants of the ensemble densities given by 
eq.(\ref{rho}) and eq.(\ref{rhob}).

As eq.(\ref{rho}) and its balanced analog in  eq.(\ref{rhob}) can describe a  wide range of models, this permits us to analyze neural networks with excitatory synapse variances not only different from the inhibitory ones but also variation among them. Additionally as the choice of any of the variances, e.g. $v_{kl} \to 0, b_{kl} \to 0$ implies the switching off or the zero connectivity between $k^{th}$ the presynaptic neuron and $l^{th}$ postsynaptic one, this enables the ensemble to mimic realistic biological networks where connections can be sparse and variable due to plasticity mechanism.  We recall here that previous studies considered  models where all excitatory distribution parameters were of same strength although with options to be different from those of inhibitory ones and therefore can not model sparsity in networks. An analysis of the above model therefore is 
an important step forward and motivates our study.
An additional benefit of using above ensembles is the potential access to choose  diagonals zero by choosing 
$v_{kk} \to 0, b_{kk}=0$.  This is useful to model realistic cases: indeed in general neurons do not  form self-synaptic connections. But if the total number of diagonals  are far less than those of the off-diagonals, choosing the former non-zero is believed to have negligible effect on the properties in large $N$-limit. This motivated 
previous studies \cite{rajan2006eigenvalue}, based on dense matrices, to consider non-zero diagonals. The above however is not expected to be valid in case of sparse connectivity.  In this work, we analyze the ensembles of both types i.e dense matrices with nonzero diagonals and sparse ones with zero diagonals.

We note that the simple form of the models used in past ensured that the square of
the radius of the circle containing the eigenvalues is $N$ times the average of the variances of the 
excitatory and inhibitory distributions. The  richness of our model however rules that out.

\subsection{Eigenvalue spectrum}

As mentioned below eq.(\ref{eq1}), the stability analysis of the network can  be pursued in terms of the eigenvalues of $J$  (also referred as modes) \cite{Rajan2010}. As $J$ is in general a real asymmetric  matrix, its eigenvalues, say $\lambda_k$, $k=1 \to N$ can be real or form complex conjugate pairs. 
The real part of a complex eigenvalue, referred later as $\text{Re}(\lambda)$,  is related to decay time of one of the activity patterns and the imaginary part, referred later as $\text{Im}(\lambda)$ is proportional to frequency of that pattern. Activation of one or more of complex modes leads to spontaneous oscillations of the network and thereby a nonlinear persistent dynamics due to their superposition.

The distribution of the eigenvalues  on the complex plane, within the bulk as well as those appearing near the edge of the spectrum, have important implications for network dynamics. In case most eigenvalues lie close to the origin,  the network then has few slow modes, leading to rapid signal decay and weak memory. By contrast, 
if eigenvalues are denser near the edge, the system retains many slow modes that 
support persistent fluctuations, high-dimensional dynamics, and more robust information processing. 
For a continuous-time neural network e.g. of type eq.(\ref{eq1}), its stability depends on the eigenvalues of $J$: the system is stable if the central bulk with complex eigenvalues  has $\text{Re}(\lambda) < 1$. 
The boundary $\text{Re}(\lambda)=1$ separating stable and  unstable spectral regions in the complex plane is referred as the line of stability. The eigenvalues to the left of the line lead to stable dynamics, those at right of the line unstable activity, and the ones on the line resulting in critical behaviour.

 For large random networks (e.g., those with identical and independent Gaussian connectivity $J_{ij} \sim {\mathcal N}(0, \sigma^2)$) without excitatory/inhibitory (E/I) structure, the eigenvalues are often distributed within a circle (e.g. Girko law), the stability depends on whether the spectral radius $S(J)$  (i.e the absolute value of the maximum eigenvalue) crosses the line of stability.
 The  network is stable for $r <1$, critical for $r=1$ and unstable/ chaotic for $ r >1$. If entire eigenvalue bulk lies left of the stability line, this indicates decay of all modes and the network settles to a fixed point i.e a quiet state. For the bulk just touching the line, the network is at critical state between stability and instability. A variation of parameters however may lead to expansion of bulk: If the bulk crosses the line with some eigenvalues developing real part $> 1$, this implies network entering in an unstable regime, permitting uncontrolled growth of activity under small perturbations.

In presence of E/I structure, the spectrum no longer remains in general a simple symmetric circle and develops, in addition to a bulk, the structures described by outliers that appears  as isolated points outside the bulk. In case the excitation and inhibition exactly cancel, the large outliers do not appear. 
In case of a slight imbalance e.g. If excitation slightly dominates, a real outlier eigenvalue appears on the positive real axis.  

The outliers i.e the eigenvalues lying outside  the main bulk  contain important 
information about network dynamics: they represent dominant global modes of activity 
where many neurons act in synchrony. A large positive outlier indicates an excitation-driven
 self-amplifying mode that pushes the network toward runaway activity resembling a pathological state e.g. seizures. A large negative 
outlier signals an inhibition-dominated mode where activity is suppressed into silence. Both cases
undermine the high-dimensional richness of bulk dynamics, collapsing the system into a
rigid, low-dimensional state dominated by a single mode. For this reason, large outliers
are considered detrimental for brain-like computation, since they drown out the diversity
of population activity needed for flexible information processing. Thus, the density profile of the bulk 
reflects how rich and long-lasting the network’s internal activity can be.

Under  perturbation, the  network may approach instability through three possible routes: (i) 
 spectral bulk crossing the line again leads to high-dimensional irregular activity, (ii) a real outlier crossing the line results in fixed-point instability and manifests as uniform activation or population mode, (iii)  a complex outlier crossing the line results in e.g. Hopf oscillations / rhythmic activity (indeed a complex outlier encode rotational modes in neural activity space, and when they destabilize, they create rhythms).
The three routes are believed to lead to brain rhythms (alpha, gamma, etc.), central pattern generators and oscillatory neural circuits

With changing parameters, the outlier may cross the stability line before the bulk does and  the network can become unstable even if bulk lies left to the line of stability. The new stability condition now requires both, the bulk stability (i.e the bulk must lie left of the line) and outlier stability (the largest real eigenvalue i.e outlier must satisfy: $Re(\lambda_{outlier})<1$). Instability occurs when either bulk  crosses the line leading to  chaos, irregular, high-dimensional activity (Sompolinsky regime) or the outlier crosses the line leading to structured instability (population mode), coherent activity, global oscillations or uniform firing in Low-dimensional dynamics and 
is often seen in population bursts and synchronized firing. Thus in  E/I networks, instability can come either from chaotic bulk modes or from structured outlier modes whichever hits the stability line first.

\section{Details of the models}

Previous studies based on simple network models indicate that the plasticity mechanisms, pharmacological effects or physiological modulators influence  synaptic variability  more significantly than the  mean synaptic strengths \cite{Rajan2010}. The mechanism is also believed to lead to dramatic predictions about the efficacy of modulatory drugs and manipulations that simultaneously strengthen and weaken subsets of synapses within the same brain region. Realistic neural networks are however often more complex e.g. sparsely connected, with type and strength of connectivity varying with spatial distance between neurones. It is therefore important to know the effect of the synaptic sparsity on the plasticity mechanism and is our basic motivation for the present study. 
To derive a clear insight about the role of sparsity in plasticity mechanisms, here  we consider two type of synaptic matrix ensembles:

(i)  {\it effectively sparse case:}  To represents the networks with densely (fully) connected neurons but  varying relative strengths of the synaptic connections, the ensemble are chosen with non-zero but decaying relative variances of the entries, 

(ii) {\it exact sparse case:} To represent the networks with missing connections among some neuron pairs, the chosen ensemble consists of both random as well as non-random entries e.g some of them exactly  zeros.

As the synaptic matrix is in general a real asymmetric matrix with a fixed fraction $f={N_E \over N_E+ N_I}$ of $N_E$ excitatory and $N_I$ inhibitory columns, its  spectral behavior is sensitive to $f$. The state with $f > 0.5$, implying the number of excitatory neurones higher than the inhibitory ones, is referred as the {\it population bias}.  Denoted by the symbol $E > I$, the population bias competes with another bias in the network, namely, {\it strength bias}, denoted by the symbol $I > E$ which in turn affects the stability and dynamics of the network. While the population bias pushes the spectrum toward positive values, the strength bias tends to make it negative; this indicates the potentially  important role played by the ratio $f$ in network dynamics.  A special case of eq.(\ref{rhob}), with same variance for excitatory columns but different from those of inhibitory columns was analyzed in \cite{rajan2006eigenvalue} for different $f$ values; In the present study, however, with our focus on the role played by the variances in network dynamics, we keep $f$ fixed and vary the excitatory/ inhibitory variances in  many possible ways.  Notwithstanding fixed $f$, each ensemble depends on at least three additional free parameters, namely, $b_e, b_i$ and system size $N$ and the nature of that dependence varies significantly. This permits us to analyze the role of varying system parameters on the spectral statistics and thereby gain insights in various aspect of network stability and plasticity mechanism. 
 For clarity of presentation,  their details are  divided in two subsections based on the dense or sparse connectivity.

\subsection{Dense (full) synaptic connectivity but with varying relative strengths}

From eq.(\ref{eq1}),  $J_{kl}$ implies the synaptic connectivity between a presynaptic neuron $k$ and postsynaptic neuron $l$, with $k, l$ referring to underlying basis indicies. Assuming that $|k-l|$  relates to the physical distance between the neuron pair, a choice of the variance $v_{kl}$ with different functional dependence on $|k-l|$ leads to the ensembles of synaptic matrices with different effectively sparsity (i.e with all matrix elements with non-zero variances but some of them relatively negligible in comparison to others).    Here we numerically analyze the ensembles with three different functional dependence of their variances (the details given below). The ensemble density $\rho(J)$  in each case is given by eq.(\ref{rho}) (for unbalanced condition) or eq(\ref{rhob}) (for balanced condition). For a clear insight in the role of variance, we keep $b_{kl}=0$ for all $k,l$ indices in each case. 

\subsubsection{\it Ensembles with uniform long range synaptic connectivity (UCE)}
This case corresponds to eq.(\ref{rho}) or eq.(\ref{rhob})  with same variance for all excitatory off-diagonals and similarly for inhibitory ones (referred hereafter as UCE for brevity),
\begin{eqnarray}
 v_{kl}= {1\over 2} \left(1+{N\over b_l} \right)^{-2}, \quad v_{kk}={1\over 2}. 
\label{ybe}
\end{eqnarray} 
with $b_l=b_e$ or $b_i$ based on whether $l$ belongs to the excitatory column or inhibitory one.

\subsubsection{{\it  Ensembles with decaying synaptic connectivity as a power law (PCE)}}
The ensemble densities in eq.(\ref{rho}) or eq.(\ref{rhob}) in this case represent a random synaptic matrix with  off-diagonal variances decaying as a power law (referred hereafter as PCE for brevity),
\begin{eqnarray}
v_{kl}={1\over 2} \left( 1+ \frac{|k-l|^2}{b_l^2} \right)^2, \quad v_{kk}={1\over 2} 
\label{ype}
\end{eqnarray}  
Assuming the same basis description as in the previous case, this case represents a network with synaptic connectivity decaying with spatial distance among pre-post neurons as a power law. In addition, the variances of excitatory columns are different from those of the inhibitory ones. With a fixed excitatory population bias ($f=0.67$) and a stability line at $Re(\lambda) = 1$, a change of $b_e$ now results in an
interplay of population imbalance and spatial structure, thereby driving the emergence of distinct spectral regimes.

\subsubsection{ {\it Ensembles with  exponentially decaying synaptic connectivity  (ECE)}}
The ensemble density now corresponds to eq.(\ref{rho}) or eq.(\ref{rhob})  with off-diagonal variances decaying as an exponential (referred hereafter as ECE for brevity),
\begin{eqnarray}
v_{kl}= {1\over 2} \, {\rm e}^{|k-l|^2/b_l^2} ,  \quad v_{kk}={1\over 2}. 
\label{yee}
\end{eqnarray}  
Here again the represented network is with the connectivity decaying with distance between basis states but the decay is now exponentially rapid. As in the previous cases, here too the variances of excitatory columns are different from those of inhibitory ones thus mimicking a realistic network more closely. 

We note that the diagonal elements in each of the three cases are typically non-zero. This however does not affect the statistics, the number of non-zero off-diagonals being $N$ times larger than the diagonals. A similar approach was used in \cite{rajan2006eigenvalue} too.

The above three cases with dense connectivity corresponds to all off-diagonals typically non-zero although the nature of their distribution  varies from constant variance  to  power law decay to exponential decay. As a typical off-diagonal matrix element is not exactly zero,  these cases can at best reflect only an {\it effectively} sparse network.  As discussed in next section, the eigenvalues distributions for the above cases  reveal a complex interplay between the $E/I$ population imbalance, the non-uniformity  of  the synaptic connectivity (variance) and the effect of balance constraint. We note that, in contrast to previous studies, the $E/I$ imbalance in three models mentioned above is caused by different relative variances and not by  their mean values. Indeed both real and complex outliers can appear due to different structures of $E$ and $I$ noises/ randomization strengths.

\subsection{\it Ensembles with sparse synaptic connectivity (SCE)}

We consider a network with its synaptic connectivity described by a real assymmetric matrix $J \in \mathbb{R}^{N \times N}$,
with its  elements  given by
\begin{equation}
J_{ij} =
\begin{cases}
+s_j \, |X_{ij}|, & |i-j| \le R,\quad i \neq j, \\[6pt]
0, & |i-j| > R,
\end{cases}
\label{sp}
\end{equation}
where the variable $s_j$ encodes the neuronal identity based on  Dale’s law $s_j =+1$, for excitatory neuron, and $-1$, for inhibitory neuron. The random variables $X_{ij}$ are independently drawn from Gaussian distributions,
$X_{ij} \sim \mathcal{N}(\mu_{E}, \sigma_{E}^2)$  and $X_{ij} \sim \mathcal{N}(\mu_{I}, \sigma_{I}^2)$ for excitatory and inhibitory cases, respectively, with mean $\mu_E = \mu_I = 0$
and population-dependent variances $\sigma_E^2 = b_e$, and $\sigma_I^2 = b_i$.

In contrast to previous cases, the above case corresponds to an ensemble of sparse synaptic matrices with many off-diagonals exactly zero.  We note that the diagonal elements in this case are exactly zero, thereby leading to a zero-trace matrix  and a fixed trace ensemble (with trace zero).  Following from eq.(\ref{sp}), all outgoing synapses from an excitatory neuron are strictly positive, whereas all outgoing synapses from an inhibitory neuron are strictly negative, enforcing Dale’s law at the level of individual neurons. The parameter $R$ defines a hard interaction range such that synaptic connections are restricted to a band of width $2R+1$ around the diagonal. Within this band, synaptic weights are random and uncorrelated, while couplings outside the band vanish identically. To understand the role of sparse connectivity along with excitatory and inhibitory strengths, we analyze the ensemble for both unbalanced and balanced conditions, each case  for three different values of $R$ i.e $R=5, 15, 60$ and for many $b_e$ values while keeping $b_i$ fixed. As described below, the variation of sparsity has significant influence on the network dynamics.

\section{Behavior of the spectrum: sensitivity to connectivity}

The stability and dynamics of neural networks are fundamentally governed by the eigenvalue spectrum of their synaptic  matrix. The randomization of the latter due to underlying complexity manifests through the fluctuations of its spectrum which in turn  depend on the matrix as well as ensemble constraints. It is therefore relevant to analyze the spectral statistics of different  synaptic matrix ensembles representing different sparsity of the neural networks. Here we  numerically investigate their spectral distribution  with primary focus on the following aspects: (i) distribution on the complex plane, (ii) the spectral density of the real part of the eigenvalues,  (iii)  spectral density of the imaginary  part of the eigenvalues, (iv) the role of synaptic strength variability in determining the spontaneous dynamics of networks.

The spectral density at a point $z=x+iy$ in the complex plane for a realization of  $N \times N$  synaptic matrix $J$ (referred often as the empirical spectral distribution) is defined by

\begin{eqnarray}
\rho_J(z) = {1\over N} \sum_{k=1}^N. \delta^2(z - \lambda_k) = {1\over N} \sum_{k=1}^N. \delta(x - x_k) \delta(y-y_k)
\label{den1}
\end{eqnarray}
Here the spectral radius, referring to the absolute value of the maximum eigenvalue, is  $S(J) = \max\left\{|\lambda_1|, \ldots, |\lambda_N|\right\}$.

An average of $\rho_J(z)$  over all realizations in the ensemble leads to ensemble averaged spectral density 
$ \langle \rho_J(z) \rangle = \int \; {\rm d}\lambda_2 \ldots {\rm d}\lambda_N \,  P(z, \lambda_2, \ldots, \lambda_N) $ with $\langle . \rangle$ implying the ensemble average. For ensembles of dense matrices e.g. those with i.i.d elements, $\rho_J(z)$ approaches $ \langle \rho_J(z) \rangle$ (referred as self averaging or ergodicity in spectral density) in limit $N \to \infty$. This however is not the case, in general, for sparse matrices. 

For later reference, we also define,  the  ensemble averaged density of the real part of the eigenvalues as 
\begin{eqnarray}
\langle \rho_{x}(x)\rangle = {1\over N} \langle \sum_{k=1}^N \delta(x- x_k) \rangle = 
\int {\rm d} y  \int  \; {\rm d}\lambda_2 \ldots {\rm d}\lambda_N P(x, \lambda_2, \ldots, \lambda_N).
\label{denr}
\end{eqnarray}
 Similarly 
the ensemble averaged density of the imaginary part of the eigenvalues can be described as 
\begin{eqnarray}
\langle \rho_{y}(y)\rangle = {1\over N} \langle \sum_{k=1}^N \delta(y- y_k) \rangle = 
\int {\rm d} x  \int  \; {\rm d}\lambda_2 \ldots {\rm d}\lambda_N P(x, \lambda_2, \ldots, \lambda_N). 
\label{deni}
\end{eqnarray}

Previous studies considered a rank 1 perturbation of a dense random matrix ensemble resulting in change of mean synaptic strength which in turn led to appearance of  outliers. The ensembles described in section III.A, however correspond to effective sparsity or exact sparsity governed by some free parameters and as shown by our numerical analysis, the outliers here appear as sparsity is varied. To understand the sparsity dependence of the spectrum,  we exactly diagonalize  each ensemble for a fixed  matrix size $N= 600$ but for many combinations of the ensemble parameters.  The ensemble size ${\mathcal M}$  i.e the number of  matrices in each ensemble is chosen  so as to give  a smooth behavior and therefore depends on the specific measure.

\subsection{ Distribution on complex plane}

As mentioned in section I, the connections in synaptic networks are subjected to variation due to plasticity mechanism. A direct comparison of the distributions 
on the spectral plane for an ensemble with increasing variances or changing sparsity type can then provide important  insights in the plasticity mechanism. Below we consider the eigenvalue distribution on the complex plane for many $b_e, b_i$ combinations but for a fixed E/I population ratio ($f=0.67$). 
As discussed blow, the eigenvalues distribution  in each case  reveals a complex interplay between the $E/I$ population imbalance, the non-uniformity  of  the synaptic connectivity (variance) and the effect of balance constraint.

\begin{figure}[H]
    \centering
       \includegraphics[width=\textwidth,height=0.70\textheight,keepaspectratio]{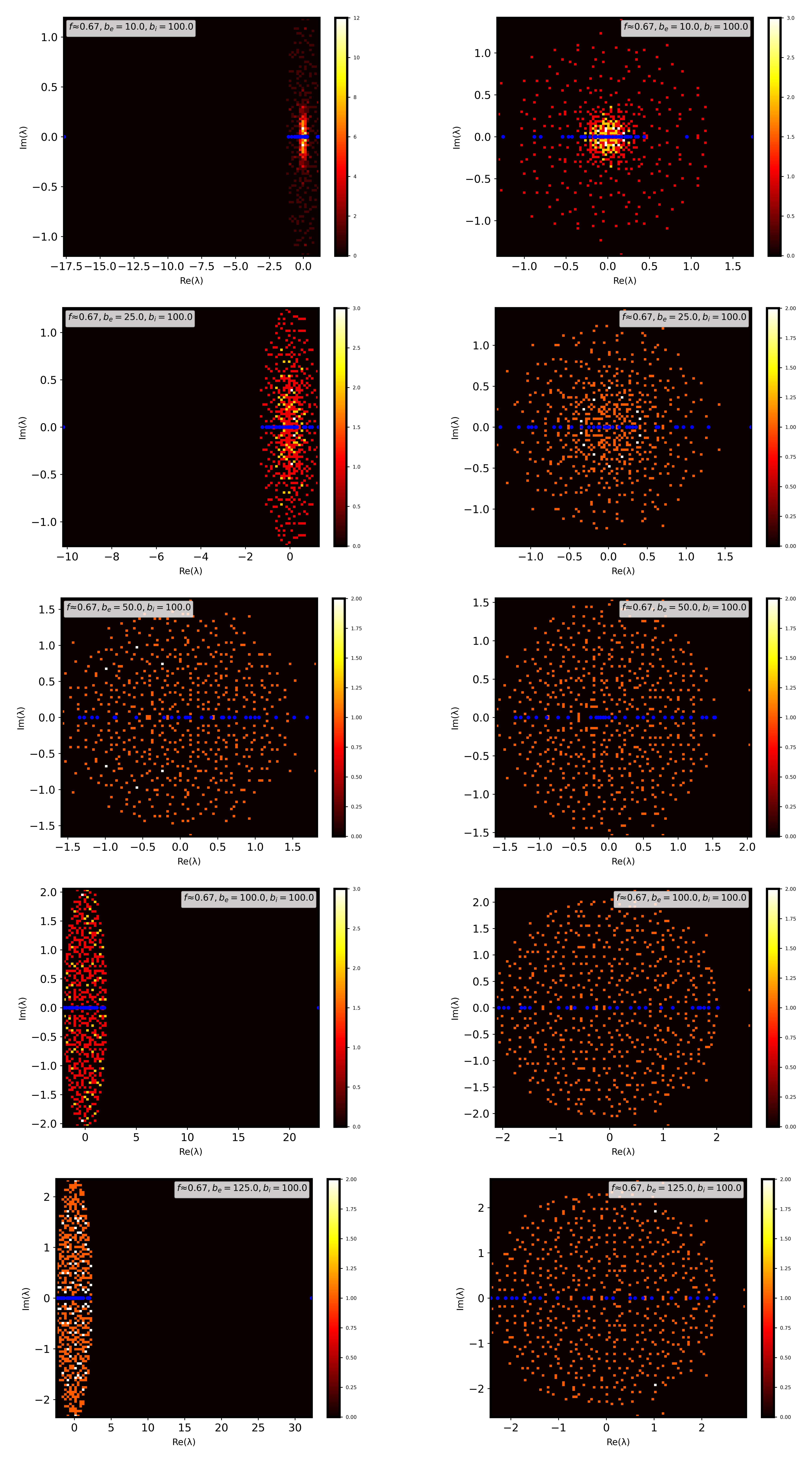}
    \caption{Eigenvalue density spectra for the UCE (constant variance) model of a random synaptic matrix ($N=600$, \red{$f=0.67$}), comparing unbalanced (left column) and balanced (right column) conditions. Blue points indicate real eigenvalues ($\text{Im}(\lambda)=0$), which determine the stability of non-oscillatory modes. The unbalanced matrices, constructed with only Dale's Law, show a dominant outlier eigenvalue, a dense cluster separated from the bulk—that destabilizes the network. In contrast, the balanced matrices have a "row-sum zero" condition imposed, which removes the outlier and centers the spectral "bulk" at the origin, stabilizing the network. The rows demonstrate the effect of varying E/I variance parameters ($b_e$, $b_i$): symmetric variances ($b_e = b_i$, top row) produce a uniform circular disk, while asymmetric variances ($b_e \neq b_i$, bottom rows) create a non-uniform disk with a high-density central zone.}
\label{dist-be}
\end{figure}

{\bf  Case(i) UCE:} With variance given by eq.(\ref{ybe}), the excitatory ($b_l \equiv b_e$) and inhibitory  strength ($b_l \equiv b_i$)  are a measure of the synaptic connectivity, with latter increasing with former. For our analysis, we begin with a large $b_i$ thereby implying the initial state of the network as a  inhibiton or {\it strength biased} state (i.e $I > E$) and increase $b_e$  while keeping  $b_i$ fixed to reach the {\it population biased} state (i.e $I < E$). 
To illustrate the effect of varying network conditions, figure \ref{dist-be} displays the spectral distribution for many $b_e$ values and for both unbalanced (left panel) as well as balanced cases (right panel).  

{\it Unbalanced condition:} The display in left column reveals a hidden "tug-of-war" between the $E>I$ population bias and the $I>E$ strength bias that leads to a transition from a  {\it quiescent state} to a {\it seizure state}. 
 In {\it rows 1} and {\it 2}, the inhibitory connections are significantly stronger ($b_i=100$) than the very weak excitatory connections ($b_e=10, b_e=25$ respectively). The strength imbalance dominates, creating a powerful net inhibition driven dynamics that manifests as a large negative real outlier ($\mathrm{Re}(\lambda) \approx -17.5$ and $-10$, respectively). This spectrum represents a quiescent state in which the network is "clamped" by overwhelming inhibition, with mean inhibitory drive silencing the network leaving the latter incapable of spontaneous activity.   As the excitatory strength increases, a significant change occurs: a critical point, seemingly marking a transition, appears in {\it row 3} ($b_e=50$) where the opposing mean-field tendencies ($E>I$ population vs. $I>E$ strength) approximately cancel and the outlier vanishes. Finally, in {\it rows 4} and {\it 5} ($b_e=100, b_e=125$), the excitatory strength is strong enough for the $E>I$ population bias to win, creating a large positive real outlier ($\text{Re}(\lambda) \approx 22$ and 32); this represents a pathological, seizure-like state of runaway excitation \cite{rajan2006eigenvalue,haider2006neocortical} (i.e a state that pathophysiologically causes an inappropriate \& excessively synchronous repetitive depolarisation of relatively large groups of neurons e.g. cortical neurons which are phylogenetically older ones as in hippocampus). We note the central bulk with complex eigenvalues still has $\text{Re}(\lambda) < 1$ and  the eigenvalues only on real line exceed the line of stability.

{\it Balanced condition:}  As the display in right column indicates, invoking  the row-sum zero constraint successfully removes the mean-field outlier for all five $(b_e, b_i)$ combinations. The network's dynamics is now governed by the randomized bulk with its spectral radius expanding as $b_e$ increases. In {\it row 1}, the weak $b_e$ ($=10$) results in a modest total variance, producing a small circular bulk that remains  left of the $\mathrm{Re}(\lambda)=1$ stability line. This is a stable randomized state, representing a healthy, functional network with complex background activity. However, as $b_e$ increases from $25$ to $125$ ({\it rows 2-5}), the increasing total variance expands the bulk's radius, causing it to cross the $\mathrm{Re}(\lambda)=1$ line. These correspond to unstable randomized states where the network's own fluctuations are self-amplifying, leading to a high-dimensional, randomized "explosion" that would saturate a real network.

\begin{figure}[H]
    \centering
  \includegraphics[width=\textwidth,height=0.70\textheight,keepaspectratio]{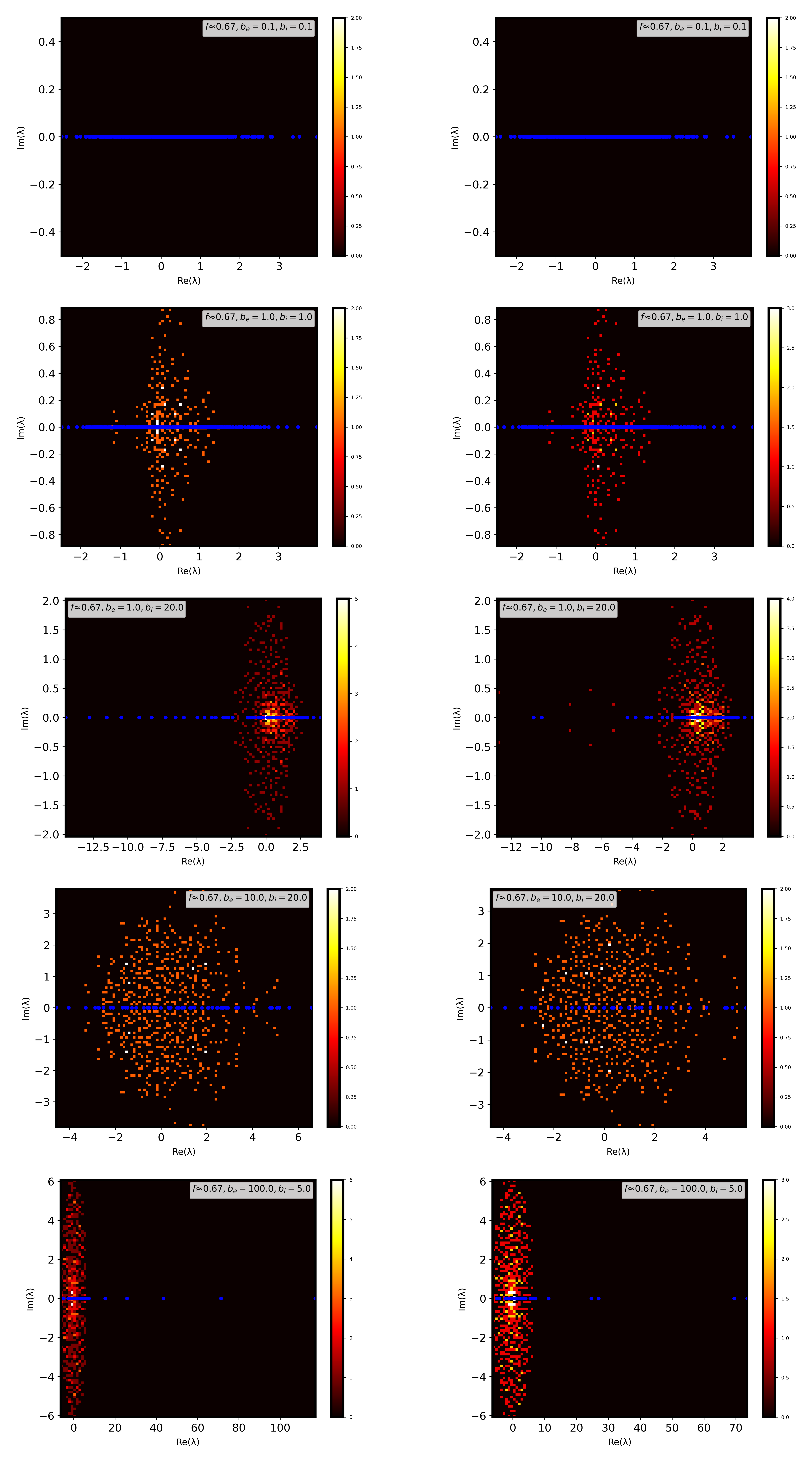}
    \caption{Eigenvalue density spectra for the PCE (Power-Law Decay) model of a synaptic matrix ($N=600, f=0.67$), comparing unbalanced (left column) and balanced (right column) conditions. The PCE model introduces spatial correlations where variance decays as a power law with distance. Unlike the uniform disks of the constant-variance (UCE) model, these spectra exhibit a pronounced condensation of eigenvalues onto the real axis, forming a characteristic "blade" or "cross" shape, particularly for smaller variance parameters (top rows). This indicates a prevalence of non-oscillatory, persistent modes ideal for memory storage. The balance condition (right column) reduces the magnitude of extreme outliers compared to the unbalanced case (left column), aiding stability, though large real eigenvalues persist in high-variance regimes (bottom rows) due to the heavy-tailed nature of the spatial connectivity. Parameters $b_e$ and $b_i$ control the spatial range of excitatory and inhibitory connections, respectively.}
  \label{dist-pe}
\end{figure}

{\bf Case (ii): PCE } Here again we consider both the unbalanced and balanced conditions for five $b_e, b_i$ combinations; the results are displayed in left and right columns, respectively, in figure \ref{dist-pe}.

{\it Unbalanced condition:} As displayed in {\it left panel}, {\it row 1}, due to weak local connectivity ($b_e = 0.1, b_i = 0.1$), the spectrum collapses entirely onto the real axis (flat line), indicating zero oscillation (due to lack of complex eigenvalues). The spectral range however extend to ${\rm Re}(\lambda) \approx 3.5$, thus exceeding the stability line ${\rm Re}(\lambda)=1$; the network is thus not quiescent but exhibits a slow, monotonic instability where activity drifts and grows without oscillating. In {\it row 2} ( $b_e = 1.0, b_i = 1.0$) as connections strengthen, the spectrum forms a "{\it blade}" shape with complex eigenvalues lying away from the real line; (a {\it blade}" shape here refers to the spectral density on the real axis that appears as a sharp edge due to no complex eigenvalues in the neighborhood). Indeed the real eigenvalues extend further past the stability line (reaching ${\rm Re}({\lambda}) \approx 3.5$). This indicates a stronger instability driven by the population bias $E > I$, now capable of supporting both growth and potential oscillations (arising from complex modes). In {\it row 3}  with $b_e = 1.0, b_i = 20.0$, a relatively stronger inhibition now overcomes local excitation and the spectrum shows massive negative real outliers stretching down to $-12$. These outliers act as a "structural brake" or clamp, pulling the system towards stability and counteracting the excitatory push. in {\it row 4}  with $b_e = 10.0, b_i = 20.0$), a critical regime appears where excitation and inhibition spatial ranges are competitive. The spectrum expands into  bulk on the complex plane (later referred as complex bulk for brevity) without the pathological outliers seen in other rows. While it crosses the stability line (${\rm Re}(\lambda) > 1$), the instability is distributed across many complex modes rather than a single structural direction. This represents a rich, high-dimensional randomized state ideal for information processing. In {\it row 5} with $b_e = 100.0, b_i = 5.0$, at extreme spatial scales, the network breaks down. The spectrum shows massive positive outliers reaching ${\rm Re}(\lambda) > 100$. This represents a catastrophic, non-oscillatory explosion where the activity hits the stability limit almost instantly.

{\it Balanced Condition:} (displayed in right panel in figure \ref{dist-pe}) Here again the "row sum zero" balance constraint  removes the single global mean-field outlier but fails to alter the fundamental structural dynamics. In the local and intermediate regimes ({\it rows} 1 \& 2), the spectra remain nearly identical to the unbalanced case, with real eigenvalues still exceeding the stability line (${\rm Re}(\lambda) > 1$). This confirms that the instability is driven by local connectivity variance, which global balancing cannot suppress. In {\it row 4}, the complex chaotic bulk is preserved and centred near the critical line, maintaining the dynamical flexibility required for learning. In {\it row 5}, the most striking failure occurs in the seizure regime. Balancing reduces the positive outliers only slightly (${\rm Re}(\lambda)$ now decreasing to $\approx 70$) but fails to bring them anywhere near the stability line (${\rm Re}(\lambda)=1$). Clearly an extreme spatial mismatch ($b_e \gg b_i$) creates an inherent structural pathology that no amount of synaptic balancing can seemingly fix. This indicates a homeostatic failure i.e an absence of a self-regulating process by which the network maintain a stable internal environment despite changes in external conditions; this stability or equilibrium, is essential for organisms to function effectively and efficiently.

\begin{figure}[H]
    \centering
    \includegraphics[width=\textwidth,height=0.70\textheight,keepaspectratio]{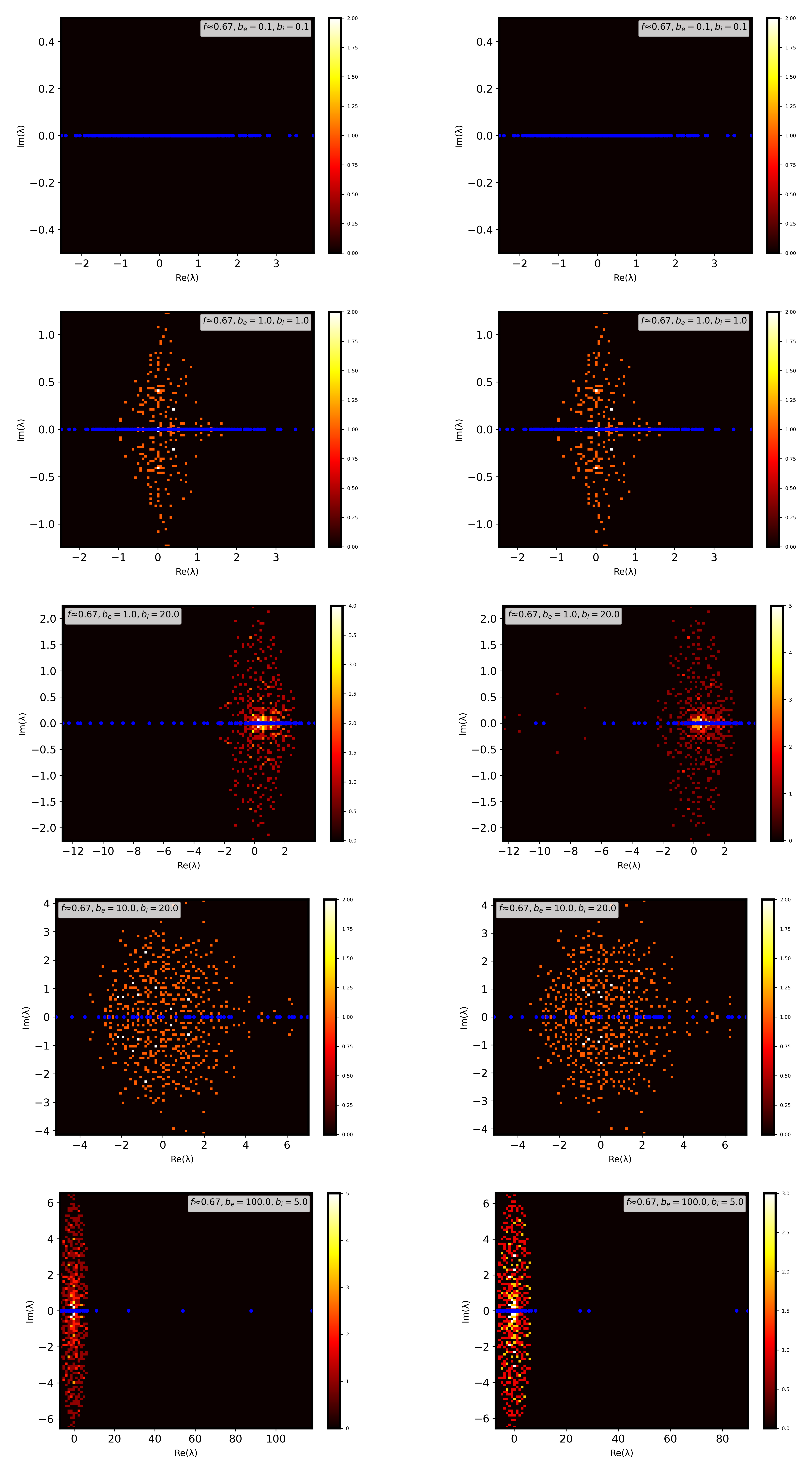}
    \caption{Eigenvalue density spectra for the ECE (Exponential Decay) model of a synaptic matrix ($N=600, f=0.67$), comparing unbalanced (left column) and balanced (right column) conditions. The ECE model introduces spatial correlations where variance decays exponentially with distance. Similar to the power-law (PCE) case, the spectra exhibit a strong condensation of eigenvalues along the real axis (the "blade" as blue points), corresponding to non-oscillatory, persistent modes. The balance condition (right column) successfully removes the largest mean-driven outlier present in the unbalanced case (left column) but fails to remove the "blade" structure or stabilize the large real eigenvalues caused by spatial correlations. This indicates that in spatially structured networks, global homeostatic balance is insufficient to suppress the amplification of spatially correlated modes.}
 \label{dist-ee}
\end{figure}

{\bf Case(iii) ECE:} In analogy with the PCE model for the network, the  model in eq.(\ref{yee}),  defined by a fast, local exponential decay, also reveals the network undergoing through  different stability/ activity phases with increasing $b_e$. Indeed, despite the steeper spatial cutoff compared to the power-law, the network’s dynamics remains governed by the connectivity range ($b_e, b_i$) relative to the stability line at ${\rm Re}(\lambda) = 1$.  Figure \ref{dist-ee} displays the distribution on the complex plane for both unbalanced as well as balanced conditions. Here again the progression of dynamical regimes mirrors the PCE model, establishing that spatial scale, rather than decay shape, is the primary driver of spectral structure.  

{\it Unbalanced condition:} For $b_e = 0.1, b_i = 0.1$, displayed in  {\it row 1} of the left panel of figure \ref{dist-ee}, the spectrum collapses onto the real axis but extends to ${\rm Re}(\lambda) \approx 4.0$. This confirms a monotonic instability identical to the PCE case, where memory traces drift rather than holding stable. In {\it row 2} ($b_e = 1.0, b_i = 1.0$), the "blade" structure again emerges, with real outliers now reaching $\approx 4.0$, thereby driving a stronger instability. A structural clamping of the spectrum occurs in {\it row 3}: with broad inhibition ($b_e=1.0, b_i = 20.0$), the "blade" of negative real outliers appears ($\approx -12$), effectively clamping the bulk dynamics and neutralizing the excitatory bias.  As excitation broadens ($b_e = 10.0, b_i=20.0$), a transition to chaotic dynamics manifests in  {\it row 4}, with the spectrum expanding into a large complex bulk extending to ${\rm Re}(\lambda) \approx 5$. This represents a robust high-dimensional chaotic state. With further increase of $b_e$ leads to a seizure state ({\it row 5}); this is caused by extreme nonlocal excitation ($b_e = 100.0, b_i=20.0$) creating massive positive outliers (${\rm Re}(\lambda) \approx 110$), replicating the catastrophic "super-hub" failure seen in the PCE model. 

{\it Balanced condition:}  The network dynamics with increasing $b_e$ but under the balanced condition is displayed in figure \ref{dist-ee}, right column. The results  reinforces the claim that the network topology overrides homeostasis. More specifically,  {\it rows 1 \& 2} indicate persistent instability, with balance condition  failing to contain the local instabilities. The spectra remain topologically identical to the unbalanced case, with real eigenvalues still exceeding the stability line (${\rm Re}(\lambda) > 1$). Further increase of $b_e$ indicates, in {\it row 4}  a functional plasticity; here the complex bulk is preserved and centered, providing the necessary dimensionality for 
rich dynamical regimes, though it remains functionally active (crossing the stability line). The display in {\it row 5} reveals homeostatic failure; as with the PCE model, here too balancing reduces the seizure outliers only marginally (to $\approx 80$), proving that the pathological instability driven by extreme spatial mismatch ($b_e \gg b_i$) is intrinsic to the anatomy and effectively immune to synaptic balancing.

{\bf Case (iv) SCE:} With $R$ as the parameter describing the sparsity of the synaptic connectivity, it is desirable to consider its role for various $b_e, b_i$ combinations. Here we consider three possible values of local connectivity, namely, $R=5, 15, 60$ for both unbalanced and balanced conditions.

\begin{figure}[H]
    \centering
    \includegraphics[width=\textwidth,height=0.70\textheight,keepaspectratio]{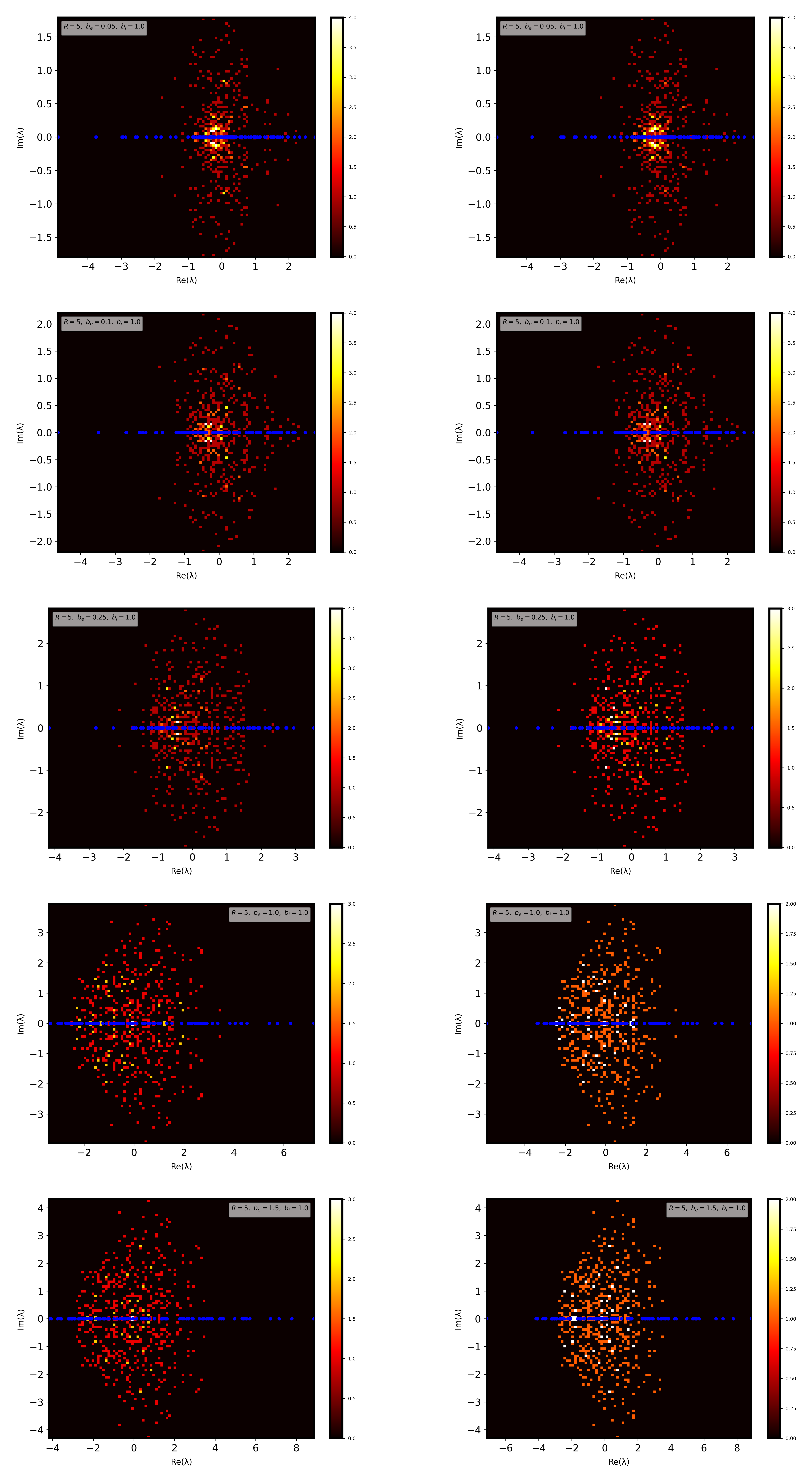}
    \caption{Eigenvalue spectra for SCE model ($N=600$, interaction range $R=5$) across increasing variance $b_e$ (Rows 1–5). Left: unbalanced condition, Right: balanced condition. In contrast to  the circular geometry for Ginibre ensemble, the spectra exhibit a robust elliptical geometry. Increasing variance ($b_e$) drives a geometric split: the complex bulk (oscillatory modes) diffuses towards the stable negative real side, while the real outliers (non-oscillatory modes) extend aggressively into the unstable positive domain ($Re(\lambda) > 0$). The persistence of this "positive blade"  under the balanced condition confirms that local topology dominates global homeostatic constraints.}
\label{dist-ae1}
\end{figure}

{\it Unbalanced condition:} The left panel of figure \ref{dist-ae1} illustrates,  for $R=5$, the spectral evolution on the complex plane with changing $b_e$ with $b_i=1$ kept fixed. Contrary to the "spectral collapse" seen in local power-law (PCE) and exponential-decay (ECE) models, a substantial spread in the imaginary dimension (reaching ${\rm Im}(\lambda) \approx \pm 3.0$ for large $b_e$) confirms that even with a short interaction range, the network creates sufficient local loops to support complex, oscillatory dynamics and enters in a critical geometric split driven by increasing variance ($b_e$).  In the low $b_e$ regime, the spectrum is characterized by a high-density core (bright yellow/white) tightly clustered near the origin. As excitation increases, this central bulk begins to diffuse significantly and enters in a critical geometric split, revealing a fundamental asymmetry in the network's stability in {\it row 4, 5}.  The complex eigenvalues (the oscillatory bulk) now tend to concentrate on the negative real side, suggesting that the oscillatory modes remain largely stable or damped. Simultaneously, and in sharp contrast, the real outliers (the tips of the ellipse) extend  towards the positive real side, reaching values as large as ${\rm Re}(\lambda) \approx +7$ in both balanced and unbalanced cases. This indicates that the network's instability is driven almost exclusively by non-oscillatory modes, while the oscillatory dynamics are effectively pushed into the stable regime. 

{\it Balanced condition:}. A comparison between the left and (right) columns in figure \ref{dist-ae1} indicates almost similar response to the unbalanced and balanced conditions for all $b_e, b_i$ combinations. This in turn 
 highlights the dominance of this sparse network topology over global homeostasis: while the balance condition succeeds in centering the bulk slightly, it fails to suppress the dangerous positive real outliers or correct the elliptical distortion. The persistence of this "positive blade/ negative bulk" structure in the balanced case confirms that the separation of stable oscillations and unstable integrators is a fundamental topological feature of the $R=5$ connectivity which global constraints cannot override.

\begin{figure}[H]
    \centering
    \includegraphics[width=\textwidth,height=0.70\textheight,keepaspectratio]{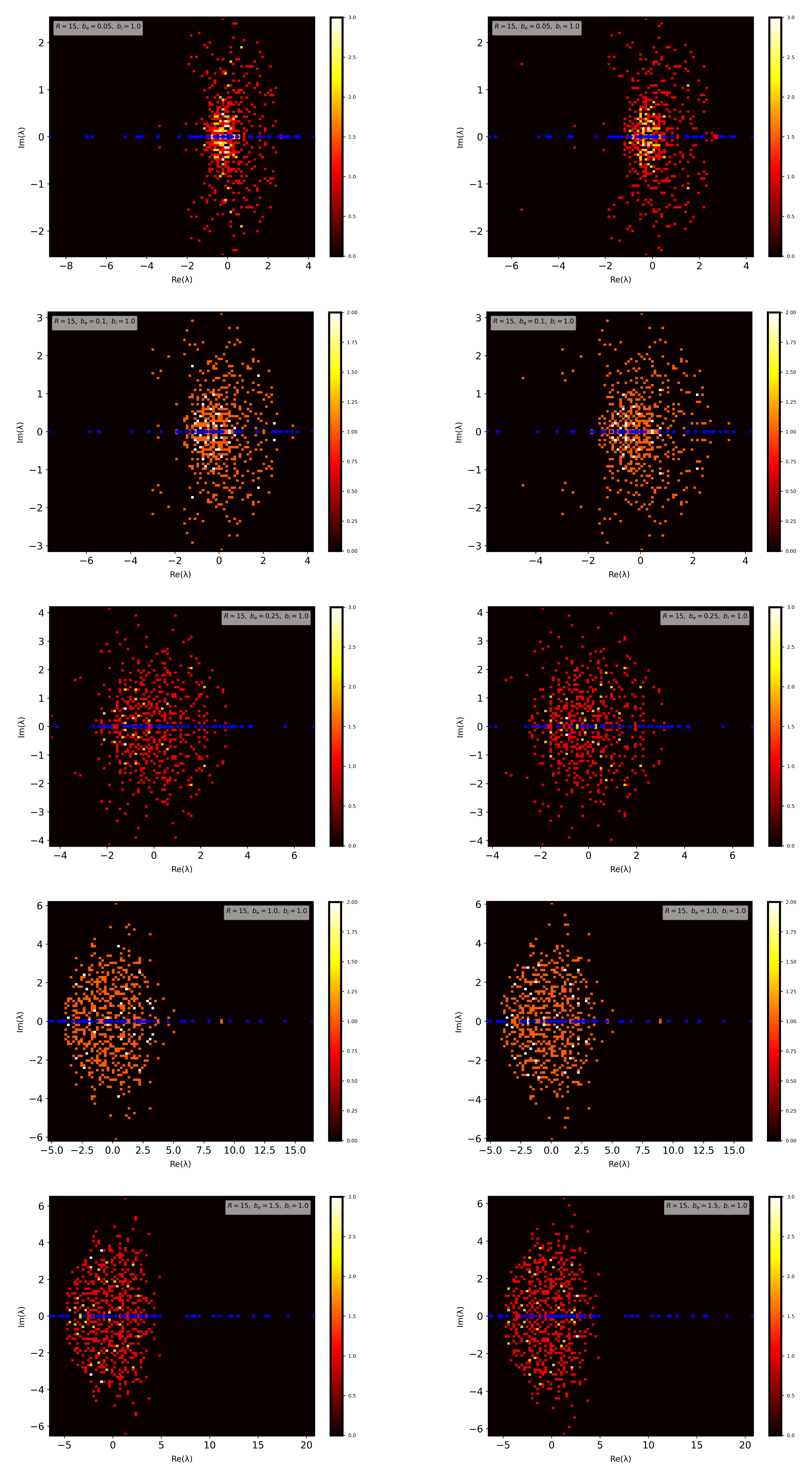}
    \caption{Eigenvalue spectra for SCE model ($N=600$, $R=15$) across increasing variance $b_e$. Left: unbalance and Right: balanced. Increasing the interaction range to $R=15$ triggers a geometric expansion, widening the imaginary spectrum compared to the $R=5$ case. However, a strong asymmetric drift persists: as variance increases (Row 5), the oscillatory bulk (complex eigenvalues) shifts towards negative real axis, while the non-oscillatory outliers (real eigenvalues) extend to extreme positive values (${\rm Re}(\lambda) \approx +15$). The persistence of this skewed "D" shape in the balanced condition confirms that intermediate spatial structure effectively resists global homeostatic stabilization.}
  \label{dist-ae2}
\end{figure}

 Figure \ref{dist-ae2} illustrates the SCE case with an intermediate interaction range $R=15$, with {\it left panel}  for the unbalanced case and  {\it right panel} for the unbalanced case.   As seen from the figure, in contrast to the low interaction regime ($R=5$), the increase in $b_e$ (implying an increase in excitatory connectivity) now triggers a significant "geometric relaxation." The small elliptical strip observed for $R=5$ now  begins to "inflate," evolving into a broader, bulbous distribution. The imaginary spectrum expands considerably (reaching ${\rm Im}(\lambda) \approx \pm 5$ in the high-variance regime), confirming that the threefold increase in synaptic range ($0.5\% \to 1.5\%$) allows for the formation of longer polysynaptic loops, thereby supporting richer and more robust oscillatory dynamics. Despite this expansion, the asymmetric density drift becomes even more pronounced. As $b_e$ increases ({\it rows 3–5}), the spectrum approaches a distinct "D-shape" topology: the latter refers to the high-density core of complex eigenvalues (the oscillatory bulk) retreating  into the negative real domain (stabilizing), while the real outliers (the "blade" shape) move deeper into the positive unstable region. 
  In {\it row 5}, the real eigenvalues reach extreme values of ${\rm Re}(\lambda) \approx +15$, completely dwarfing the imaginary width. This confirms that while increasing the range $R$ allows for more oscillation, the system's instability is still overwhelmingly driven by the non-oscillatory, spatially localized modes that form the "blade" shape.  Here again the effect of the balance condition  remains marginal. While visually the bulk appears slightly more compact than in the unbalanced case, the fundamental topological distortion persists. The balanced network still exhibits the massive positive real extension and the suppression of the circular symmetry. This indicates that at $R=15$, the network is still firmly in the "spatially dominated" regime. The local clustering is strong enough to decouple the "hubs" from the global mean field, preventing the balanced constraint from enforcing the circular law or containing the unstable outliers.

\begin{figure}[H]
    \centering
    \includegraphics[width=\textwidth,height=0.70\textheight,keepaspectratio]{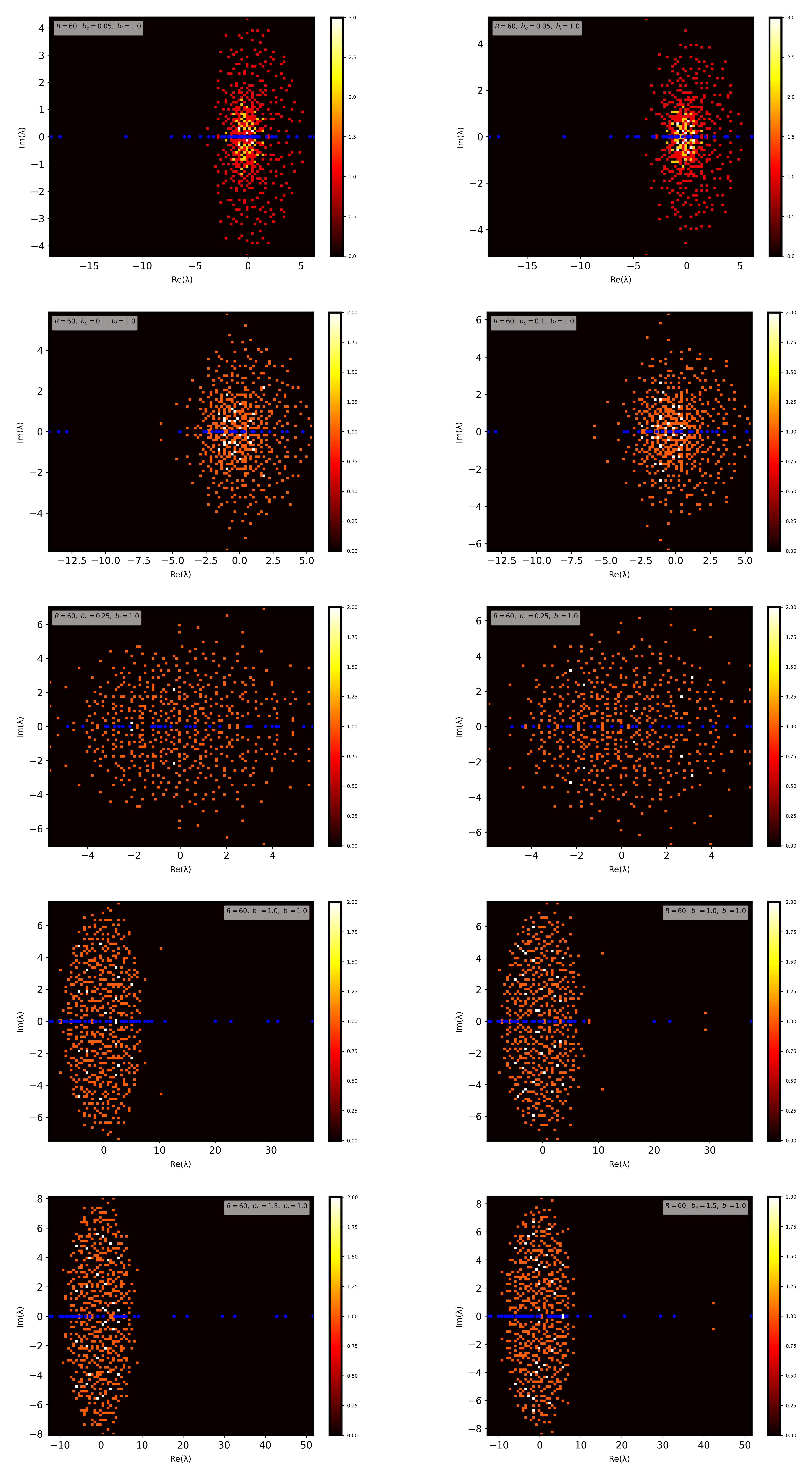}
    \caption{Eigenvalue spectra for SCE model ($N=600$, $R=60$). The left column displays the unbalanced condition, while the right column displays the balanced condition. (Rows 1–2): Low variance results in a compact, stable bulk. (Row 3 - Transition State): The spectrum is optimally centered. This stability emerges because the opposing mean-field forces—population Bias ($E > I$) and strength bias ($I > E$) approximately cancel, naturally balancing the system without artificial constraints. (Rows 4–5, Seizure State): As variance increases, the forces decouple. The complex bulk retreats to the stable negative side, while the real outliers explode to extreme positive values ($Re(\lambda) \approx +50$). The persistence of these outliers in the Balanced condition (Right) confirms that global constraints cannot fix the instability caused by uncompensated mean-field forces.}
\label{dist-ae3}
\end{figure}

The quest to understand the spectral behaviour under long-range connectivity motivates us to analyze SCE with $R=60$ (representing $6\%$ of the network). Figure  \ref{dist-ae3} illustrates the $b_e$ driven eigenvalues dynamics in this case. As figure indicates,  this regime marks the transition towards global mean-field behavior, (with rigid "strip" or "D"-shapes seen at local scales for $R=15$  relaxing into a massive "global ellipse")
yet the topological distortions remain.  The imaginary spectrum expands further reaching ${\rm Im}(\lambda) \approx \pm 10$), confirming that the network now supports a rich diversity of complex oscillatory modes comparable to random networks. This expansion however  is accompanied by an extreme asymmetric drift. In the high-variance regime ({\it row 5}), the separation between stability, represented by bulk, and instability, represented by outliers, is maximized. While the high-density core of complex eigenvalues forming the bulk is pushed deep into the negative real domain (${\rm Re}(\lambda) \approx -10$), the real outliers move in the opposite direction, reaching large positive values i.e ${\rm Re}(\lambda) \approx +50$. This confirms that in large-scale spatial networks, the "blade" of non-oscillatory modes decouples completely from the bulk, driving a brute-force instability that dwarfs the oscillatory dynamics. The display in ({\it row 3}) reveals a critical "transition state" in which the spectrum appears centred and balanced; this state arises due to opposing mean-field forces i.e population bias and strength bias effectively neutralize each other (and not due to zero row sum constraint). This in turn stabilises the spectrum where the bulk stays near the origin without requiring additional constraints; indeed this is the only regime where the spatial network naturally behaves like a stable dynamical system. Another noteworthy aspect persisting even for large $R$ is the negligible difference between the unbalanced  and balanced  cases. As in the unbalanced case, the large  positive outliers persist (${\rm Re}(\lambda) > 40$) in {\it row 5}. This indicates  that the balancing modelled by {\it row-sum zero constraints} cannot replicate the delicate "force cancellation" seen in {\it row 3}. The network topology of $R=60$ is strong enough that unless the mean-field forces ($N$ vs $b$) are tuned to cancel exactly, the network will inevitably drift into a non-oscillatory seizure state.

\subsection{Ensemble averaged Density of the real and imaginary spectral parts}

While the distribution on the complex plane provides a qualitative insight  about the eigenvalue behavior on the complex plane,  it is desirable to display the density in a way that could give  better quantitative insights. For this purpose, we numerically analyze the densities of the real   and  imaginary parts of the eigenvalues, given by eq.(\ref{denr})) and eq.(\ref{deni}), separately (latter referred as real part density and imaginary part density for brevity); the matrix dimensions used in the analysis are kept same as  in section IV.A. However, while a single matrix realization is used for the spectral analysis on the complex plane, we numerically analyze an ensemble of 10 matrices for the real and imaginary part distributions to improve the statistics optimally.

\begin{figure}[H] 
\centering
    \includegraphics[height=0.55\textheight,keepaspectratio]{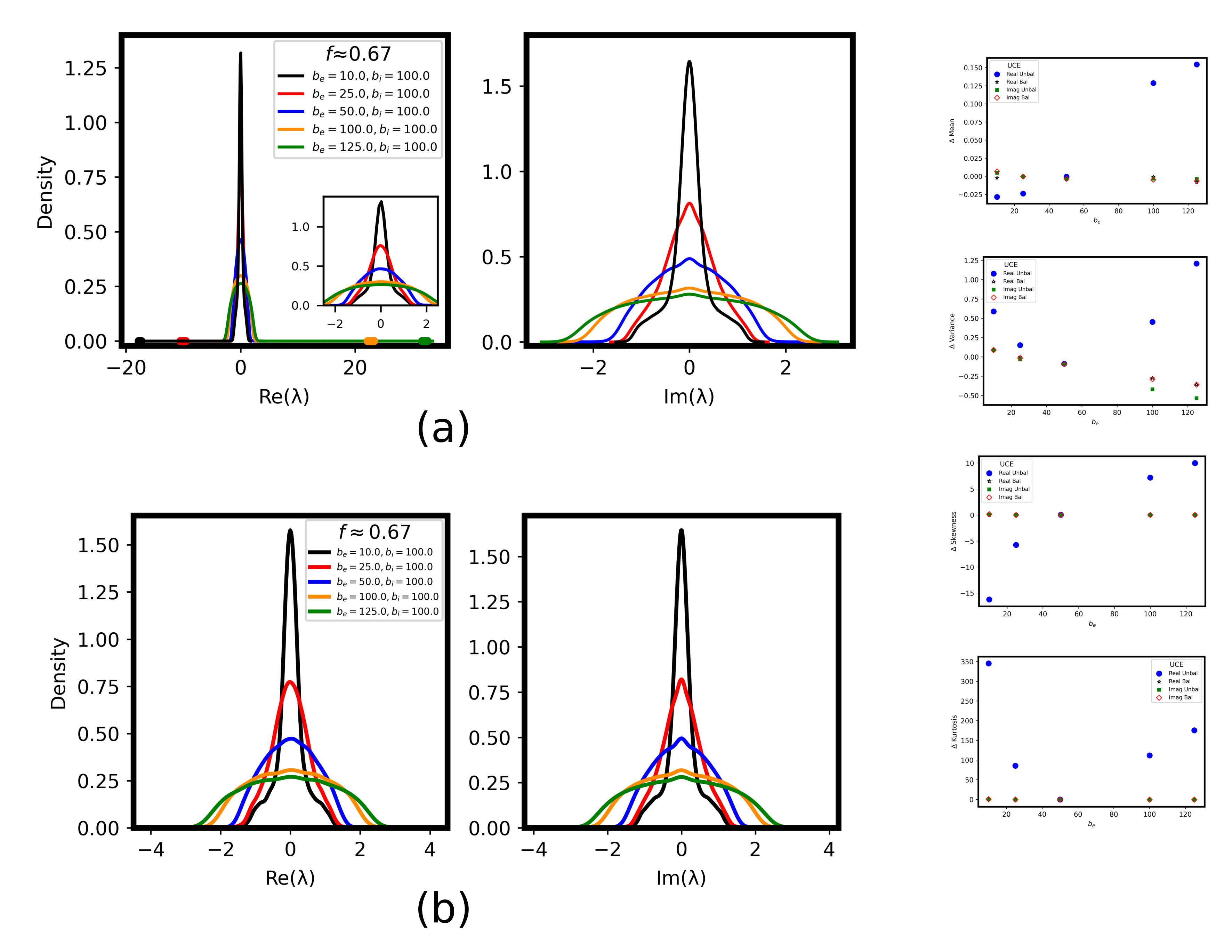}
    
\vspace{0.2cm}
\caption{Probability density of Real and Imaginary eigenvalues for the UCE (constant Variance) model across varying excitatory variance parameters $b_e$ (with fixed $b_i=100$). In the unbalanced case (left panel, (a)), the real eigenvalue density reveals the mechanism of instability: while the main bulk remains near zero, isolated outlier modes (indicated by colored dots) appear far to the left (quiescent state for low $b_e$) or far to the right (pathological state for high $b_e$), driven by the net mean synaptic drive. In the balanced case (right panel, (b)), the row-sum constraint removes these outliers, collapsing the spectrum into a single unimodal distribution centered at zero. The nearly identical shape of the Real and Imaginary densities confirms the rotational symmetry of the random matrix spectrum. As $b_e$ increases (blue $\to$ green), the distribution widens, signifying an expanding spectral radius and a transition from stable to unstable chaos. The figures in the third column display the deviations of mean, variance, skewness and kurtosis of the empirical distribution from eq.(\ref{eq:skew_t_pdf})}
 \label{den-be}
\end{figure}

{\bf  (i) UCE:}  The ensemble averaged densities of the real and imaginary parts of the eigenvalues in this case are displayed in figure \ref{den-be},  left and right panels, respectively. The part(a) and (b) of the figure depict the spectral density behaviour  of the ensemble without and with row sum zero constraint.

Figure \ref{den-be}(a), (left panel), depicting the ensemble averaged density of $\text{Re}(\lambda)$  in the unbalanced case reconfirms a competition between the tendency to remain in the "spectrum bulk",  centred around $\lambda=0$, and appearance of the isolated outlier modes (marked by single coloured dots on the x-axis in  figure  \ref{den-be}(a), left panel) which indicate the instability of the network. Based on the excitatory variance strength, here again two different states of the network emerge: for low excitatory variance (e.g., $b_e = 10, 25$), a  {\it quiescent} state appears.  The dominance of  inhibition in the network is now captured by the outliers marked by blue and red dots located far to the left ($\text{Re}(\lambda) \approx -18, -10$).  As excitatory variance increases (e.g., $b_e = 100, 125$), the balance shifts with network now approaching a  pathological state. The outliers marked by the orange and green dots appear far to the right ($\text{Re}(\lambda) \approx 22, 32$), signifying a runaway regime with excitatory drive overwhelming the system and leading to saturation. The widening of the central bulk with increasing  variance, seen in figure \ref{dist-be} (left panel) is now indicated by the crossover from the narrow blue curve to the broad green curve  but the network dynamics is effectively enslaved by the dominant mean-field outliers.

Figure \ref{den-be}(a) (right panel), depicts  the imaginary part density of the eigenvalues $\text{Im}(\lambda)$ in the unbalanced case. A broadening with increasing $b_e$ can be seen here again, thereby indicating emergence  of new oscillatory frequencies.  The lack of outliers in the figure indicate availability of only finite frequencies.

As illustrated in  figure \ref{den-be} (b) (left and right panel), imposition of the  row-sum zero constraint on the network fundamentally transforms the spectral landscape. The most important change is a complete disappearance of the outlier dots; the entire probability distribution is now confined to the central bulk. The distribution also now tends to stabilise: the real part density becomes a unimodal bell curve centred exactly at zero for all parameter sets. This confirms that balancing eliminates the pathological mean-field drive, allowing the network's dynamics to be governed solely by its fluctuation driven dynamics.

In contrast to part (a), a comparison of the left and right sub-panels of part(b) reveals the analogies in the shapes and widths of real and imaginary part densities. This  indicates that the balanced network possesses no structural bias, it is equally sensitive to decay (real axis) and oscillations (imaginary axis). The overlay of curves clearly demonstrates the expansion of the randomized bulk. As the excitatory variance increases from $b_e=10$ 
  (blue) to $b_e=125$ (green), the density curve flattens and widens significantly. While the narrow blue peak represents a stable, low-activity state,  the broad green curve indicates a transition to unstable chaos where the tails of the distribution extend beyond the stability limit ($\text{Re}(\lambda) > 1$).

Furthermore, the structure of the spectral bulk  reflects the role of the relative strengths of the underlying parameters. In the highly asymmetric variance case ({\it row 1}, $b_e=10, b_i=100$), the eigenvalue density is non-uniform and peaked at the centre. In the symmetric variance case ({\it row 4}, $b_e=100, b_i=100$), the density approaches uniformity, recovering the  circle law. The UCE model with no spatial dependence is thus dynamically limited to quiescent or randomized (either stable or unstable) states. It notably cannot produce the specialised, non-chaotic "integrator" or "selective persistent" (cross) states, strongly implying that those more functional dynamic regimes are an emergent property of the distance-dependent spatial structure in PCE and ECE models discussed later.

\begin{figure}[H] 
  \centering
    \includegraphics[height=0.55\textheight,keepaspectratio]{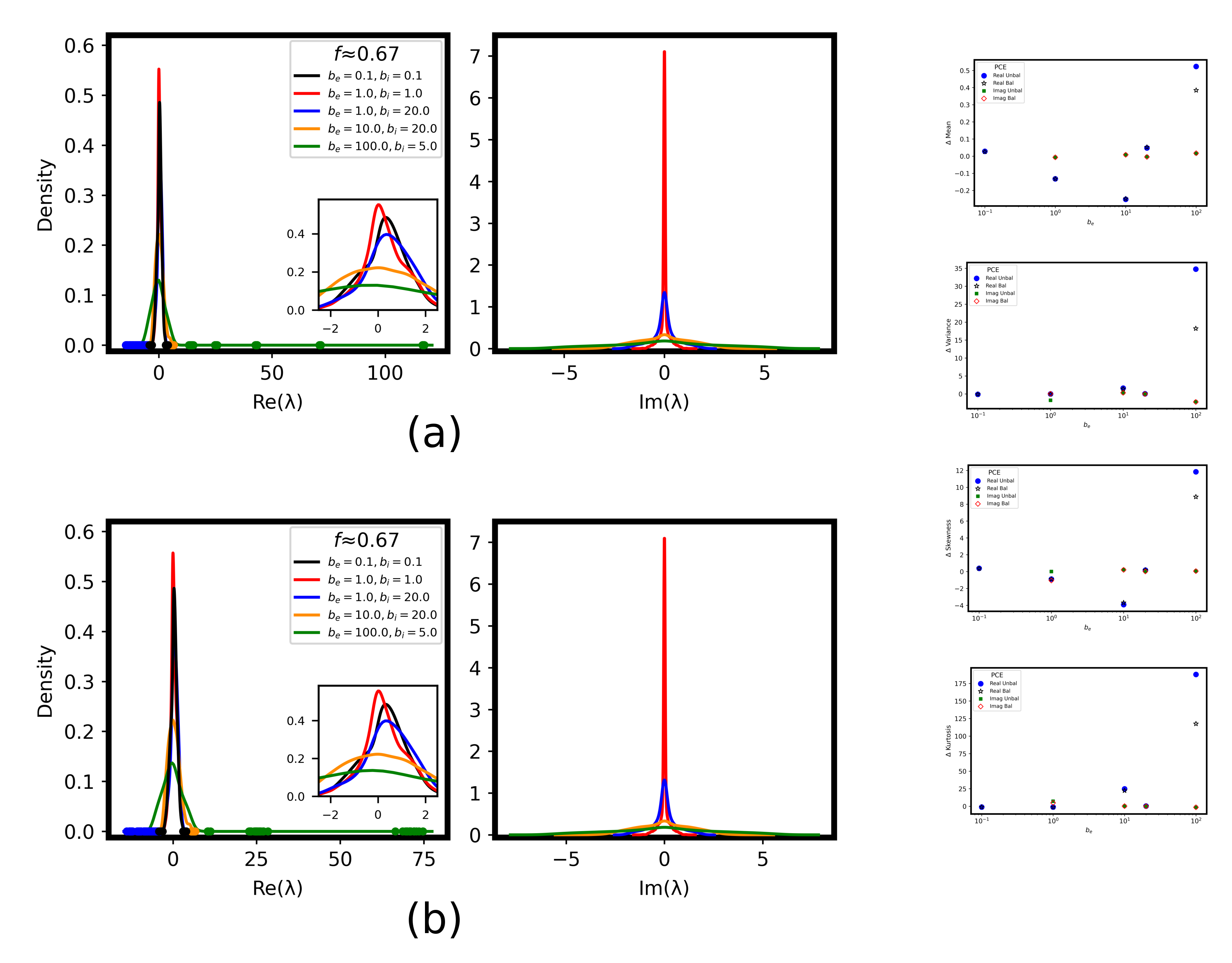}   

 \vspace{0.2cm}
    \caption{Probability density of the Real and Imaginary parts of the eigenvalues for the spatially structured PCE (Power-Law) model. In Figure \ref{den-pe}(a) (unbalanced case), the real density reveals heavy-tailed distributions driven by structural hubs. A massive tail extends far to the left (negative cross) for the inhibitory-dominated case ($b_e=1.0, b_i=20.0$, blue curve), representing a clamped state. Conversely, the tail extends far to the right (positive cross) for high excitatory variance ($b_e=100$, green curve), representing a pathological state.The peak of the real density shifts in the direction opposite to the tail (e.g., shifting negative when the tail is positive, and positive when the tail is negative). In figure \ref{den-pe}(b) (balanced plots) a distinct asymmetry is visible. For low excitatory variance ($b_e=0.1$), the black curve is absent in the imaginary plot, signifying a complete spectral collapse onto the real axis and a transition to pure integrator dynamics. The deviations of mean, variance, skewness and kurtosis of the empirical distribution from eq.(\ref{eq:skew_t_pdf}) are displayed in right column.}
 \label{den-pe}
 \end{figure}

{\bf PCE:}  Characterized by structural asymmetry, spectral collapse in local regimes, and a compensatory spectral shift, the real and imaginary part  density profiles for the PCE model reveal a fundamental departure from the  UCE model.  Here the density of  $\text{Re}(\lambda)$ in unbalanced case (figure \ref{den-pe}(a), left panel) has a distinct structural feature: while the probability mass is dominated by the central bulk, the distribution is heavily skewed by the power-law spatial decay of the synaptic connection strengths. In contrast to the isolated outliers appearing in UCE model, the density in this case is continuous and heavy-tailed; the latter is 
indeed a manifestation of the "blade" type structure seen in figure \ref{dist-pe}. For lower excitatory variance $b_e < b_i$ (e.g., $b_e=1.0$, blue curves), the distribution has a massive negative tail extending far to the left, thereby indicating a dominance of the structural hubs due to inhibition that clamp the network into a silent, non-responsive state. This also confirms that the heavy-tailed power-law connectivity of the synapses creates strong local clusters that drive the network into a runaway state much more aggressively than 
UCE connectivity. A critical feature worth notice here is the behavior of the spectral peak (the mode): despite the system having a positive trace (the sum of eigenvalues $> 0$), the peak of the green curve shifts towards the negative side. This "recoil" away from the positive outlier indicates a conservation mechanism: the network manages to compensate for the extreme instability of the hubs (the tail) by forcing the bulk of the neurones into a faster-decaying regime (negative shift). This effectively segregates the "seizing" hubs from the "quiet" majority. For high variance ($b_e = 100$, green curve), the  tail of the distribution now  extends to large positive values i.e $\text{Re}(\lambda) \approx 110$). Although  few in number (low density), these outlier modes act as structural hubs that drive local runaway excitation.

Imposing the balanced condition  re-centers the distributions near zero (figure \ref{den-pe}(b)), but crucially, it fails to restore symmetry or complete stability. A comparison of the densities of the real (left panel) and imaginary eigenvalue parts (right panel) now reveals a striking asymmetry. The real density is broad with heavy positive tails, while the imaginary density is sharp and narrow. This "symmetry breaking" indicates that the network is no longer chaotic and has a structural bias towards non-oscillatory (real) dynamics.

For a low excitatory variance ($b_e = 0.1$, black curve), the imaginary density  reveals a critical feature, namely, $\text{Im}(\lambda) = 0$ for all modes,  i.e a complete collapse of the spectrum entirely onto the real axis. Referred as the "silent" spectrum, the network now operates in a strictly non-oscillatory state with 1D dynamics. As connectivity broadens slightly ($b_e = 1.0$, red curve), a sharp peak appears but the system remains overwhelmingly non-oscillatory. For higher variance (green curve), the real density retains a significant positive tail even after balancing. This demonstrates that global balancing cannot eliminate the local structural instability caused by power-law clustering. As in the unbalanced case, the "recoil" appear here too: the peak of the real part density shifts in the direction opposite to the tail (e.g., shifting towards negative value  when the tail is positive and positive when the tail is negative). This "recoil" maintains the conservation of the trace, allowing the bulk of the network to adjust its excitability to compensate for the extreme behavior of the hubs (tails).

{\bf  ECE:} 
 Visibly distinct from the UCE case, the real and imaginary part density plots for this case 
largely mirror the PCE (Power-Law) behavior, suggesting thereby  that the emergence of structural asymmetry, spectral collapse, and compensatory shifts is a shared property of the distance-dependent connectivity. In  the left panel of figure \ref{den-pe}(a) for unbalanced case, the density of $\text{Re}(\lambda)$ exhibits the manifestation of continuous "blade" structure as in the PCE model, driven by the spatial clustering of connections.

In contrast to  UCE model where outliers appear as isolated dots separate from the bulk, the ECE model shows a heavy-tailed distribution connecting the bulk to the outliers. While the tail extends far into the negative domain for low variance ($b_e=1.0, b_i=20.0$, blue) indicating structurally driven quiescence, it spreads massively into the unstable domain (i.e $\text{Re}(\lambda) > 100$) for high variance ($b_e=100$; green). This confirms that exponential clusters, like power-law hubs, creates local feedback loops that drive runaway excitation. As in the PCE case, the compensatory shift of the maximum density is visible here again: the peak of the green curve shifts to the negative side to counterbalance the massive positive tail, conserving the trace. This confirms that the "quieting of the bulk" to accommodate unstable hubs is a general feature of spatial networks.  The real part density (figure \ref{den-pe}(a), left panel) remains broad with heavy tails, while the imaginary density (figure \ref{den-pe}(a), right sub-panel) is sharp. This asymmetry proves that the ECE model, like the PCE model, has a structural bias towards stable, non-oscillatory modes (the cross) rather than the chaotic mixing seen in random networks. For the weak local connectivity $b_e=0.1$, the Imaginary part density effectively collapses to zero again leading to a seizure state. This reconfirms that the weak excitatory connectivity forces the network into a strictly non-oscillatory state regime with 1D dynamics. 

\begin{figure}[H] 
  \centering
    \includegraphics[height=0.55\textheight,keepaspectratio]{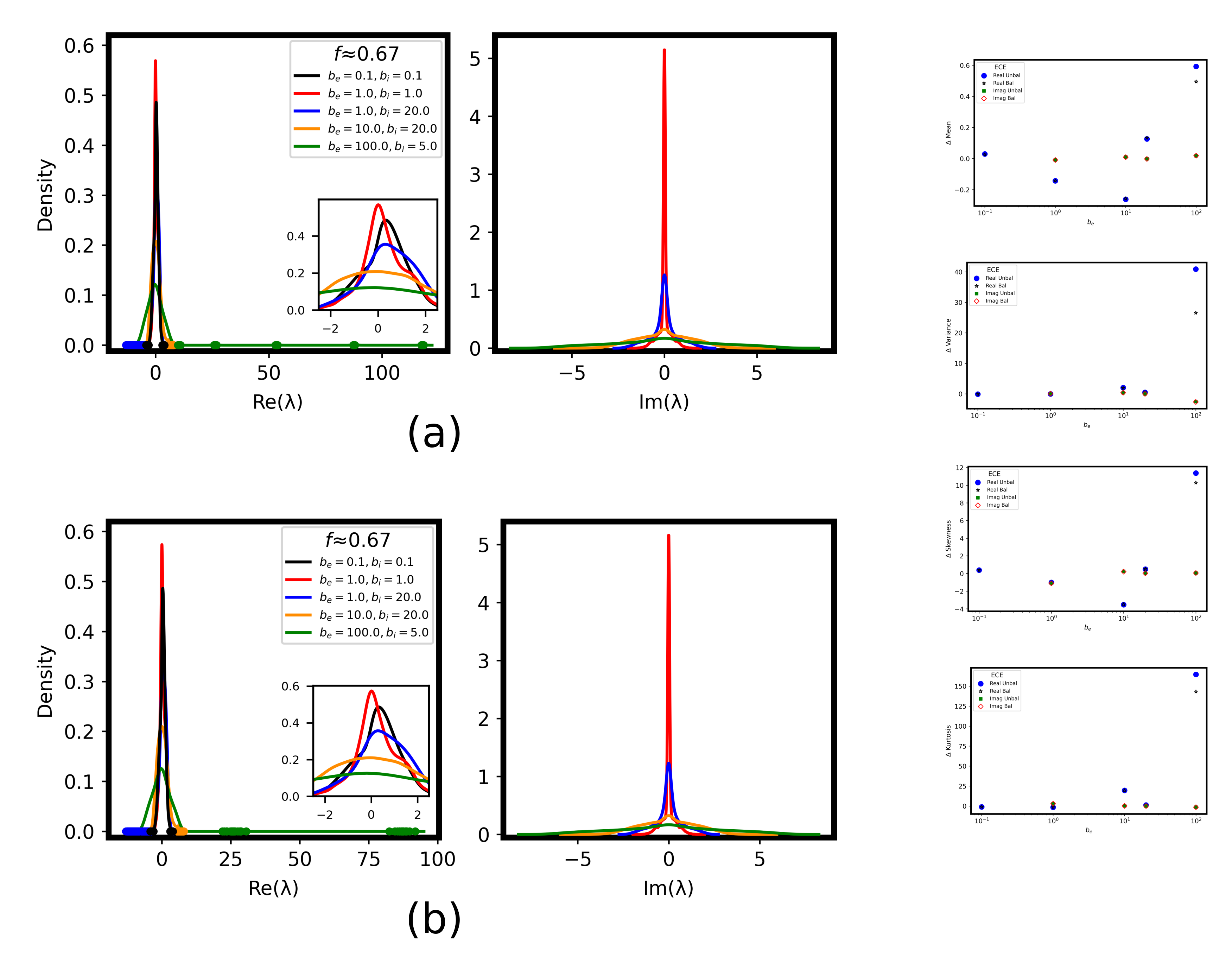}     

\vspace{0.2cm}
    
    \caption{Probability density of the Real and Imaginary parts of the eigenvalues for the ECE (Exponential Decay) model ($N = 600$). (a) Unbalanced Condition: The density profiles mirror the PCE model, establishing a universal dynamical signature for spatially structured networks. (b) Balanced Condition: The imposed row-sum constraint fails to restore rotational symmetry. Across both panels, the model exhibits structural asymmetry (broad Realbladevs. sharp Imaginary peak), spectral collapse to strictly 1D dynamics for local connectivity ($b_e = 0.1$, collapsed Imaginary density), and spectral recoil (peak shifting opposite to outliers). This confirms that local spatial correlations, whether power-law or exponential, robustly generate persistent non-oscillatory modes ("the cross") that resist global homeostatic balancing. The deviations of mean, variance, skewness and kurtosis of the empirical distribution from eq.(\ref{eq:skew_t_pdf}) are displayed in right column.}
 \label{den-ee}
\end{figure}

As the display in figure  \ref{den-ee}(b) indicates, imposing the balance condition in ECE model again fails to restore the rotational symmetry characteristic of the UCE model. The persistence of the positive tail in the real density (green curve) under balancing condition  proves that the local instability caused by exponential clustering is robust to global homeostatic control. The "hubs" created by short-range exponential decay are strong enough to defy the global row-sum constraint.

\begin{figure}[H]
\centering
 \includegraphics[height=0.55\textheight,keepaspectratio]{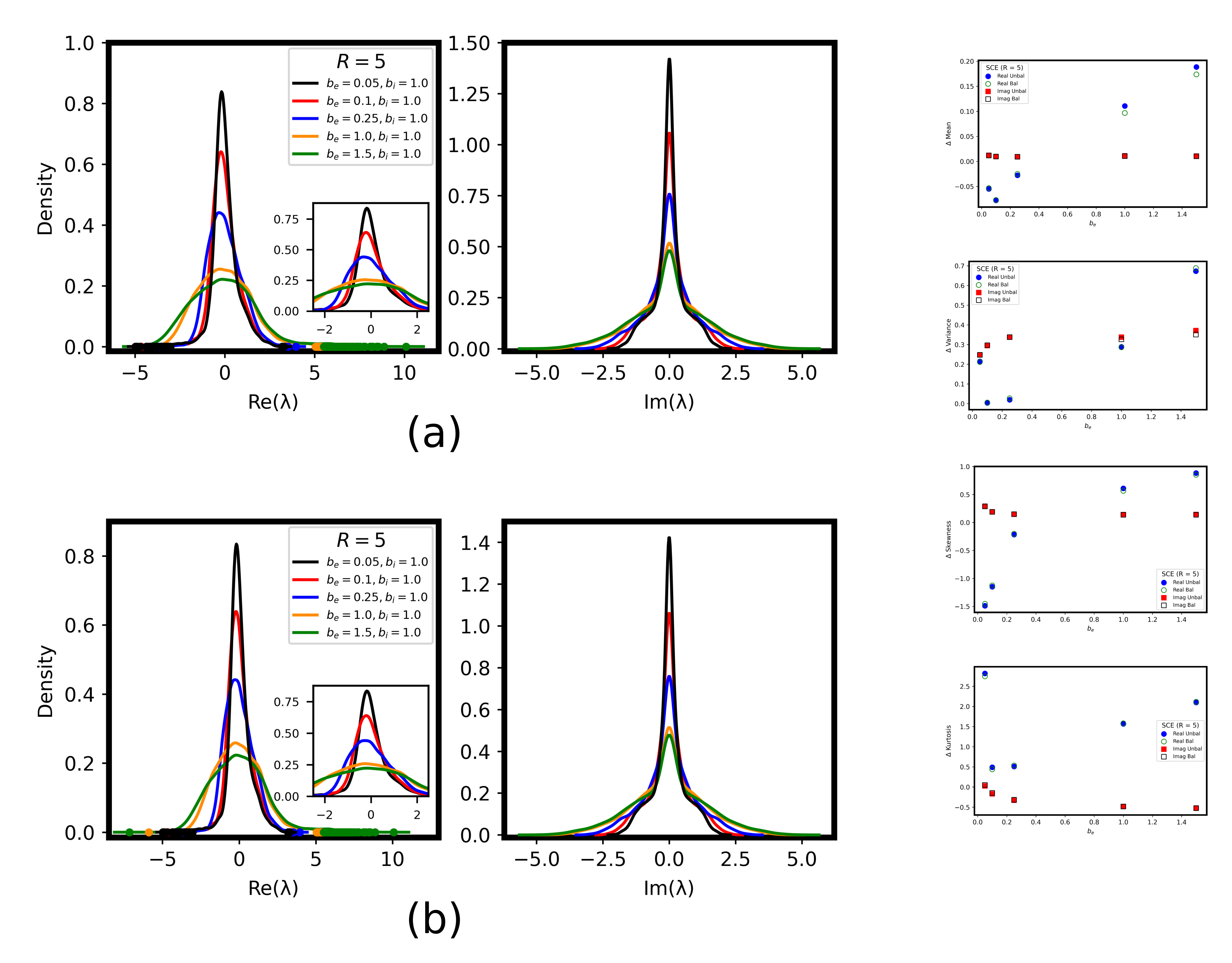}
 
 \vspace{0.2cm}
\caption{ Probability density of real and imaginary eigenvalues for the SCE model ($N=600, R=5$). (a) unbalanced Condition. (b) balanced Condition. The distributions exhibit sharp, mirroring of the "local regime" of PCE/ ECE models. The real density (left) is consistently broader than the imaginary density (right), quantifying the structural bias towards non-oscillatory dynamics. Increasing variance (red $\to$ green) broadens the tails, driving the "real blade" instability. The striking similarity between (a) and (b) confirms that global balancing fails to suppress the heavy tails or restore spectral symmetry, a universal failure mode of locally connected networks. The deviations of mean, variance, skewness and kurtosis of the empirical distribution from eq.(\ref{eq:skew_t_pdf}) are displayed in right column.}
  \label{den-ae1}
\end{figure}

\begin{figure}[H]
\centering
 \includegraphics[height=0.55\textheight,keepaspectratio]{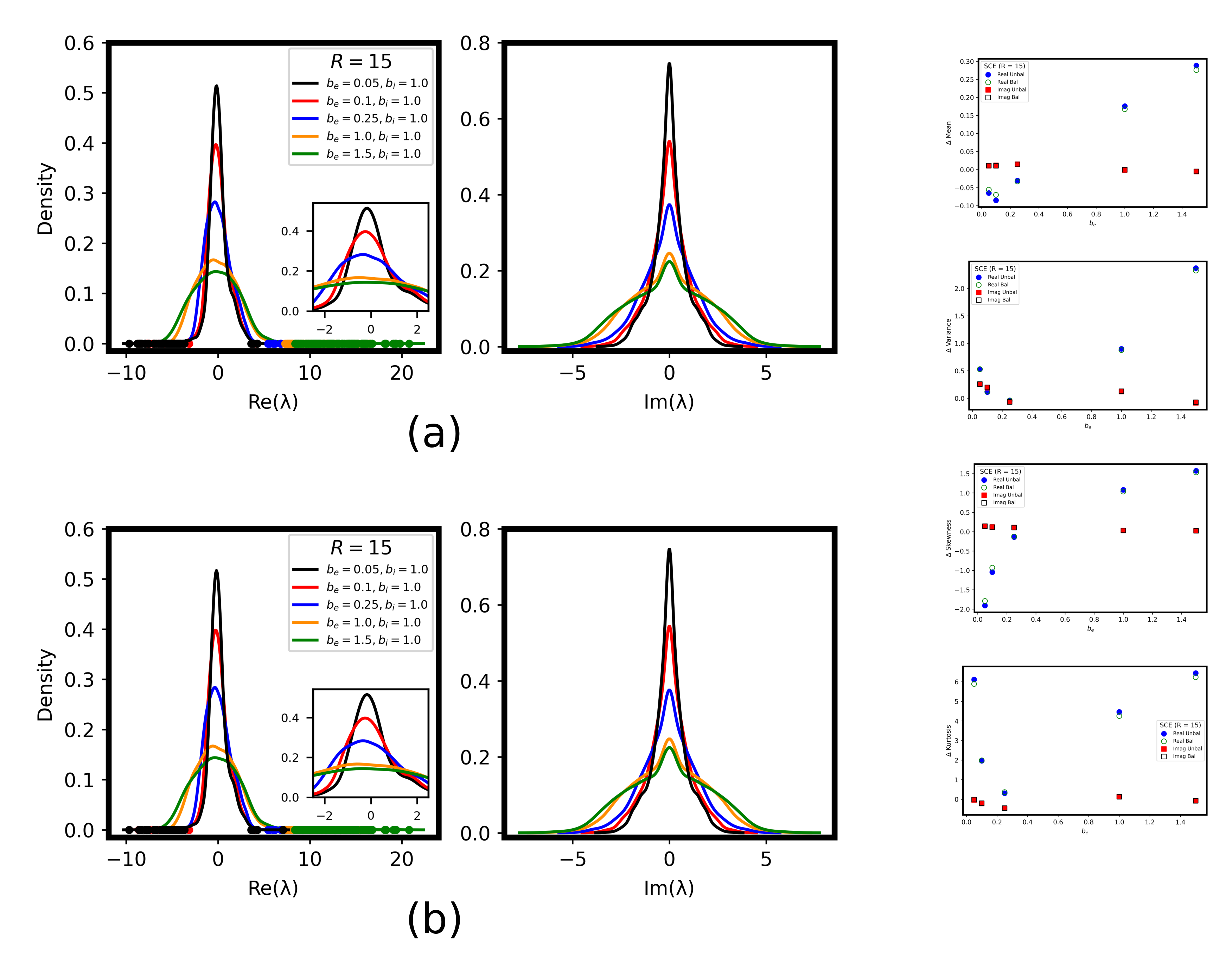}
 
\vspace{0.2cm} 
   \caption{  Probability density of real and imaginary eigenvalues for SCE model ($N=600, R=15$). (a) unbalanced condition. (b) balanced condition. The anisotropy is maximized, with real density (left) tails reaching $\text{Re}(\lambda) \approx 20$, dwarfing the imaginary density (right) ($\pm 7$). As variance increases (green curve), the positive tail explodes, forcing the peak of the distribution to shift negative (compensatory shift) to conserve the trace. The balanced condition fails to eliminate this behavior, preserving both the extreme tails and the compensatory peak shift, confirming that the network remains functionally unbalanced despite the global constraint. The deviations of mean, variance, skewness and kurtosis of the empirical distribution from eq.(\ref{eq:skew_t_pdf}) are displayed in right column.}
 \label{den-ae2}
\end{figure}

\begin{figure}[H]
 \centering
 \includegraphics[height=0.55\textheight,keepaspectratio]{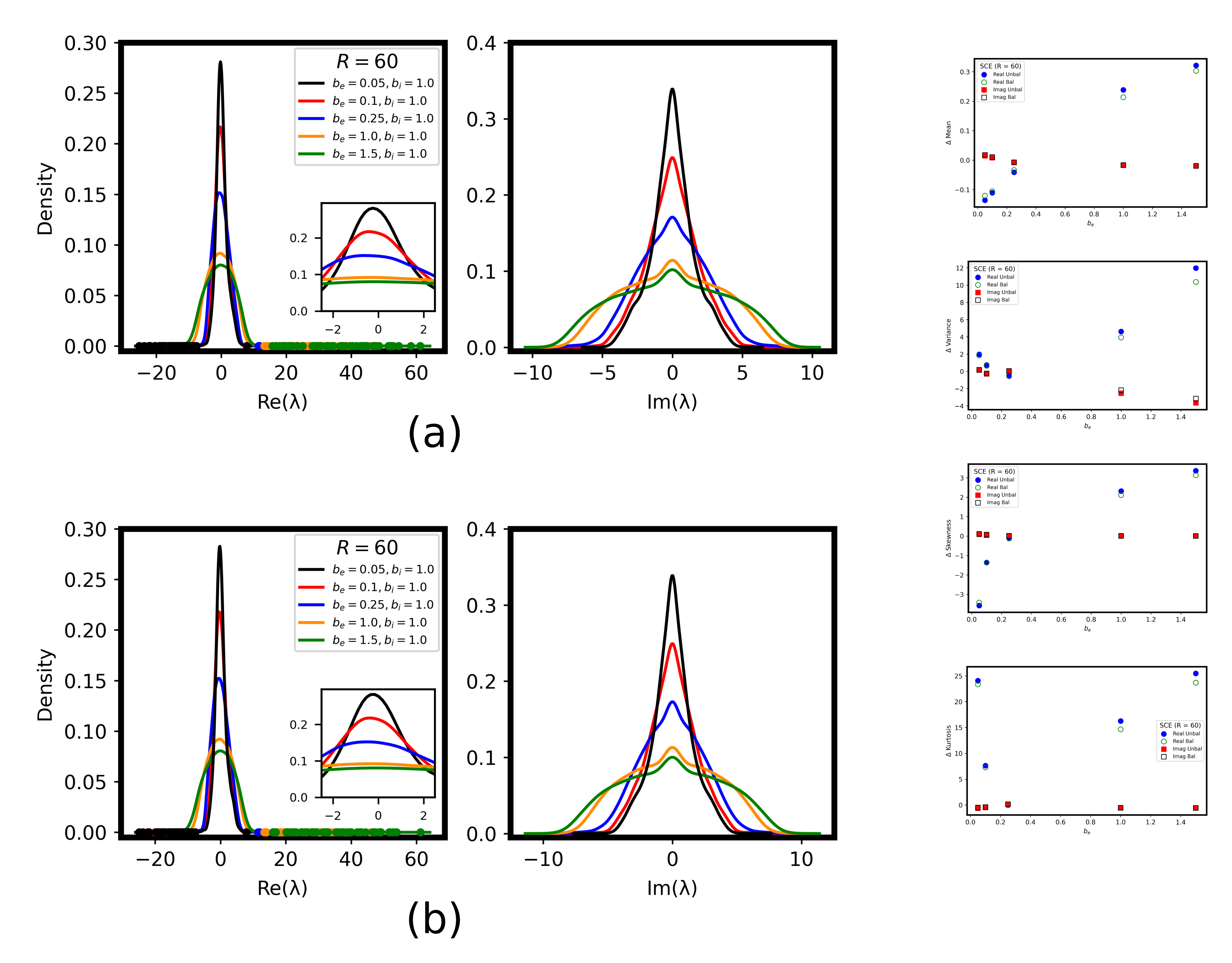}
 
 \vspace{0.2cm}
    \caption{Probability density of Real and Imaginary eigenvalues for SCE model ($N=600, R=60$). (a) unbalanced condition. (b) balanced condition. The distributions highlight the "geometric recovery" of the bulk. The Imaginary Density expands to $\pm 10$, significantly broader than the $R=5$ case, indicating the bulk is recovering a 2D elliptical geometry. (compensatory shift): As variance increases (green curve), the positive tail explodes to $\text{Re}(\lambda) \approx 60$. To conserve the trace, the peak of the distribution shifts visibly to the negative side, sacrificing bulk stability to accommodate the outliers. The balanced condition fails to correct this, preserving the shifted peak and the extreme tails. The deviations of mean, variance, skewness and kurtosis of the empirical distribution from eq.(\ref{eq:skew_t_pdf}) are displayed in right column.}
 \label{den-ae3}
\end{figure}

{\bf SCE:}  To gain quantitative insights in sparse networks behavior, we now analyze the behaviour of $\text{Re}(\lambda)$ and $\text{Im}(\lambda)$  separately for SCE, again for three $R$, many $b_e$ values and for both unbalanced as well as balanced conditions. The results are displayed in figures  \ref{den-ae1},  \ref{den-ae2},  \ref{den-ae3} for $R=5, 15, 60$ respectively, with part (a) of each figure illustrating the real  and imaginary part density for unbalanced case and part (b) for the balanced case. We recall here that the synaptic matrix in this case is a zero-trace matrix.

 As displayed in the left panel of figure \ref{den-ae1}(a), the real part spectral density in the unbalanced case for $R=5$ again indicates the dominance of structural locality over mean-field randomness. Visually, the distributions are more sharply peaked  than a Gaussian, indicating a {\it quiescence state} with high concentration of modes near zero.  A comparison of  the real and imaginary part  densities however reveals a critical asymmetry, the former  is consistently broader and possesses heavier tails than the latter;  this behavior is clearly in contrast to effective sparse cases e.g. PCE, ECE.   Indicating consistency with the approach to "elliptical" geometry seen in figure \ref{dist-ae1} ({\it left panel}), the network's phase space is now skewed, favouring non-oscillatory decay and growth over complex oscillations. As $b_e$ increases (from $b_e=0.1$ to $b_e=1.5$), the central peak flattens and the tails widen significantly. The $b_e=1.5$ case exhibits the broadest shoulders in the real domain, corresponding to the "real cross" of outliers extending into the unstable region. The broadening is far less pronounced in the imaginary domain, confirming that the instability is driven by real, non-oscillatory modes. 

Imposing the balanced  condition  for $R=5$ (figure \ref{den-ae1}(b)) highlights the failure of global homeostasis in strongly spatial networks. In contrast  to UCE model where balancing drastically transformed the landscape (removing outliers and symmetrising the bulk), here the density profiles with and without balancing   are visually analogous. The real density remains broader than the imaginary density, preserving the structural anisotropy. The heavy tails observed in the unbalanced case persist in the balanced case, confirming that the "real cross" is a topological feature of the $R=5$ band, not a mean-field artifact. Consequently, the balanced network remains susceptible to the same stationary instabilities as the unbalanced one, as the local correlations prevent the global constraint from compacting the spectrum. In addition, the density profile in this case strikingly mirrors the "local regime" of the PCE and ECE models ($b_e=0.1$). In all three ensembles, the restriction of connectivity forces a breakage of rotational symmetry, resulting in a Universality of real-mode dominance. Just as in the Power-Law case, the SCE model demonstrates that short-range interactions inherently favor a "stiff" dynamical landscape (real $>$ imaginary). However, a key difference remains, while the local PCE and ECE models exhibit near-total spectral collapse (vanishing imaginary density), SCE for $R=5$ retains a non-negligible imaginary bandwidth. This suggests that the "hard cutoff" topology  of SCE, while stiff, is far more permissive of local oscillations than the "hub-dominated" power-law topology of PCE and exponential one of ECE.

 Figure \ref{den-ae2} depicts the eigenvalue parts densities for $R=15$. In the unbalanced  case (figure \ref{den-ae2} (a)), the density profiles reveal an amplification of the structural anisotropy, driven by the emergence of "hub-like" instabilities. A direct comparison between the real part ({\it left}) and ({\it right}) panels exposes a massive scale discrepancy. The real part density  extends significantly into the unstable domain with tails reaching $\text{Re}(\lambda) \approx 20$ for large $b_e$ case (green curve). In contrast, the imaginary density  remains relatively confined ($\pm 7$). This suggests that the network is far more capable of supporting explosive non-oscillatory growth than it is of sustaining rapid complex oscillations. As $b_e$ increases (green curve), we observe the critical mechanism of trace conservation. The massive extension of the positive tail forces a visible reconfiguration of the bulk. The compensatory shift is clearly visible, the peak of the green curve shifts to the negative side to counterbalance the massive positive tail, conserving the trace. This confirms that the "quieting of the bulk" (stabilization of the majority) to accommodate the explosive instability of the "real "blade"" is a general feature of spatial networks, mirroring the behavior seen in the PCE/ ECE transition regimes. The balanced Condition (Figure \ref{den-ae2} (b)) further underscores the impotence of global constraints in the face of intermediate-scale topology. Just as in the $R=5$ case, the density profiles for the balanced network are virtually indistinguishable from the unbalanced ones. The heavy tails of the real density persist with identical magnitude ($\text{Re}(\lambda) \approx 20$). Crucially, balancing fails to restore rotational symmetry. The real distribution remains significantly broader than the Imaginary one, and the "compensatory shift" of the peak remains necessary. This confirms that the "row sum zero" rule cannot correct the fundamental topological bias that favors real-mode instability.

Figure \ref{den-ae3}  presents the eigenvalue parts densities for the long-range interaction regime ($R=60$). 
As can be seen from the {\it left panel} of both parts (a) and (b) of the figure, the real part density, $\text{Re}(\lambda)$, exhibits, for both cases, a substantial flattening and broadening, serving as the statistical signature of a spectral bulk expanding deep into the complex plane. Notably, the peak density at $\lambda=0$ drops to its minimum compared to the $R=5$ and $R=15$ cases; this signifies that eigenvalues are no longer confined to a rigid local "blade" structure but are distributed more uniformly across a wide area of the unstable region ($Re(\lambda) > 1$). Correspondingly, the imaginary part density, $\text{Im}(\lambda)$, displays its maximum width, indicating a rich diversity of  modes that drive fast, high-frequency, and high-dimensional oscillatory dynamics at the network level. Despite this recovery of the bulk geometry, the stability profile remains dominated by the explosive growth of real outliers. In the high-variance regime (green curve, $b_e=1.5$), the positive tail extends to extreme values of $\text{Re}(\lambda) \approx 60$. The most critical feature unique to this high-$R$ regime is the visible reconfiguration of the bulk to satisfy trace conservation of the ensemble. In contrast to $R=5$ case where the peak remained centred, the peak of the green curve shifts visibly to the negative side to counterbalance the large positive tail. This confirms that in large-scale spatial networks, the majority of modes (the bulk) are forced into a deeper inhibitory state solely to accommodate the runaway instability of the "real blade". With the density profiles for the balanced and unbalanced cases appearing  qualitatively analogous, this again indicates the "row-sum zero"  hypothesis for balance condition as unnecessary. More elaborately,  even in the global interaction limit, the imposition of the row-sum zero constraint fails to suppress the heavy tails or correct the compensatory peak shift, proving that the network remains functionally unbalanced despite the global constraint.

\subsection{Functional forms of the spectral densities}

In absence of the availability of an analytical formulation for  the ensemble averaged probability densities of the real and imaginary eigenvalue parts, we seek to gain insights in the functional form through various statistical fits to the numerical data displayed in Figure \ref{den-be} (a, b). For the symmetric densities, such as the imaginary parts or systems with small $b_e$, the results are consistent with a symmetric Student's t-distribution,

\begin{equation}
P\left(\frac{x-\mu}{\sigma}; \nu\right)  = \frac{\Gamma \left(\frac{\nu+1}{2} \right)}{\sqrt{\pi \nu \sigma^2} \Gamma \left(\frac{\nu}{2} \right)} \left(1 + \frac{(x-\mu)^2}{\nu \sigma^2} \right)^{-\frac{\nu+1}{2}}
\label{std}
\end{equation}

where the parameter $\nu$ varies with the system configuration. In cases where the empirical spectra exhibit significant asymmetry particularly for the real part densities, the spectral bulk is modeled using a Skew-Student-t distribution. This density, which allows for the simultaneous modeling of heavy tails and skewness, is defined by perturbing the symmetry of the Student-t density  \cite{Azzalini2003} as follows,

\begin{equation}
f(x; \mu, \sigma, \nu, \alpha) = \frac{2}{\sigma} \; P\left(\frac{x-\mu}{\sigma}; \nu\right) T\left(\alpha \frac{x-\mu}{\sigma} \sqrt{\frac{\nu+1}{\nu + \left(\frac{x-\mu}{\sigma}\right)^2}}; \nu+1\right)
\label{eq:skew_t_pdf}
\end{equation}
Here $T(x; \nu+1)$ is the cumulative distribution function (CDF) of a Student-t distribution for $x$ with $\nu+1$ degrees of freedom. The parameter $\alpha$ serves as the shape parameter regulating the skewness. For large $b_e$, a deviation near the tail is expected due to large real outliers that escape the statistical bulk.

We recall that the standard maximum likelihood estimation (MLE) for heavy-tailed distributions is inherently sensitive to extreme values. In models such as UCE, PCE and ECE, a small fraction of eigenvalues sit far outside the spectral bulk and can disproportionately dominate the log likelihood function. This "outlier pull" often prevents the optimization algorithm from accurately resolving the central shape of the distribution, potentially leading to parameter estimates that prioritize the outliers at the expense of the bulk density. Furthermore, theoretical studies have noted that the likelihood function for Student-t type models can become unbounded if the degrees of freedom ($\nu$) are allowed to span the entire range $(0, \infty)$ in the presence of extreme outliers. To mitigate these effects and focus the analysis on the continuous spectral bulk, a 1\% trimming procedure was implemented. This was achieved by excluding the extreme 0.5\% of the eigenvalues from both the top and bottom of the distribution. This selective removal of individual outliers allows the fitting algorithm to focus exclusively on the primary eigenvalue cluster. By isolating these extreme points, we ensure that the fitted parameters: $\mu$ , $\sigma$ , $\nu$ , and $\alpha$ accurately reflect the statistical behavior of the majority of synaptic interactions.
The comparison between the spectral moments obtained from the data and from the fitted parameters $\mu, \sigma, \nu, \alpha$ are presented through the moment-difference plots for the UCE, PCE, ECE, and SCE ensembles. The corresponding moment-difference figures are shown in the right-most columns of Figure \ref{den-be} for UCE, Figure \ref{den-pe} for PCE, Figure \ref{den-ee} for ECE, and Figure \ref{den-ae1}--\ref{den-ae3} for the SCE cases with $R=5,15,$ and $60$, respectively. These plots display the deviations in the mean, variance, skewness, and kurtosis between the data and the corresponding fitted distributions for both the real and imaginary parts of the spectrum. 
For small $b_e$, the deviations remain relatively small, indicating that the spectral density profiles are reasonably well described by eq.(\ref{std}), namely the Student’s-$t$ distribution (and its skewed extension where applicable). However, for larger $b_e$, noticeable deviations appear, particularly in the higher moments such as skewness and kurtosis. These deviations mainly arise from the increasing asymmetry and heavy-tail behavior of the spectral distributions, especially near the tail regions. An near analogy of numerical data with skew-student-t distribution demonstrates that the real part of the eigenvalues of synaptic matrices across UCE, PCE, ECE, and SCE models are fundamentally governed by a heavy-tailed, asymmetric statistical law. The high degree of alignment between the empirical and theoretical first two moments (mean and variance) confirms that the location and spread of the eigenvalue density are well captured by the location-scale parameters of the skew-student-t distribution.
A significant divergence is  observed however in higher order moments, particularly the kurtosis in the real part of unbalanced network. While the empirical data frequently exhibits extreme kurtosis, the theoretical fit converges to much lower values. This mismatch is however not a failure of the  predicted functional form for the density but rather a reflection of the physical distinction between the spectral bulk and isolated outliers.

\section{Conclusion}
We investigated the eigenvalue dynamics of the synaptic connectivity matrices to understand how the underlying strength and type of the randomness, arising due to complexity of the brain and varying from one brain to another, affects their stability and transient dynamics. Based on a comparative study of  different synaptic matrix ensembles with multivariate Gaussian probability density, we find  that the nature and strength of randomized synaptic connectivity is not merely a signature of network complexity but a fundamental driver of distinct dynamical regimes. Indeed,  based on the type of  synaptic connections sparsity, effective or exact, different spectral density profiles may appear on the complex plane. At very low strengths of synaptic connectivity, the bulk spectrum effectively collapses to positive real axis  irrespective of type of its randomness. The network enters,  in this regime,   into an almost non-oscillatory state with its dynamics confined to a 1D manifold of monotonic drift that is fundamentally different from the oscillatory chaos of random networks.  With increasing connection strengths, the model based features manifest into the spectrum. For example, a most defining feature of the networks with spatially sensitive, low strength  connectivity  is the emergence of a "blade" type structure formed by a heavy-tailed distribution of real eigenvalues with almost no complex eigenvalues in the vicinity; the "blade"  breaks the rotational symmetry observed in random networks. This feature seems to be a characteristic of the networks with  distance-dependent randomized connectivity, irrespective of whether the  connectivity strength decays as a power-law (scale-free) or exponentially (scale-dependent) and consistently generates the non-oscillatory backbone, segregating it from the oscillatory bulk. While a "blade" also appears in case of long range uniform connectivity models with unbalanced constraints, the real eigenvalues extend far into the positive real axis and the complex eigenvalues  now have a non-zero  but  almost zero real part and cluster around the imaginary axis. This "blade" type structure is however absent for long range uniform connectivity models with balanced constraints.  At intermediate strengths, the "blade"  may grow into an almost rotationally symmetric or elliptic bulk, implying  the network cannot distinguish between integrating a signal (real) and oscillating (complex).  In addition, except for those with uniform connectivity and with balanced conditions, an increase of synaptic strength again leads to breaking of elliptical/ spherical structure in all models considered here. Indeed while the presence or absence of balanced condition has significant impact on network models with long range uniform synaptic connectivity, it does not seem to affect those with spatially sensitive connectivity.

A fundamental question in neuroscience is how the brain maintains a critical transition state, a dynamical regime poised between quiescence and runaway saturation. Is this "balance" a form of dynamic stability? Such a state is essential for maintaining the representational capacity required for plasticity and learning. Contradicting the conventional view that the critical state is achieved simply through global homeostatic balancing (i.e row-sum zero constraint imposed at a single matrix level), our results indicate the transition to be controlled by the  type and strength of the synaptic variability and more generally by the ensemble type that reflects the underlying complexity of the network. While  imposing a global balance condition in the random UCE model  universally forces the system into a stable regime regardless of the underlying variance, effectively masking the pathological mean-field drive, it fails to ensure stability in spatially structured networks (PCE and ECE). As observed in the high-variance regimes of these cases, the 'seizure' behavior persists even after the balance condition is applied (indicated by "blade" structure with large outliers). This indicates that for the real brain, which is inherently spatial,  the "row sum zero" constraint  is not necessarily the one  that leads to a critical transition state. The latter is achieved only when the opposing mean-field forces—specifically, the population imbalance ($E > I$) and the strength asymmetry ($I > E$) cancel each other out naturally (as seen in  PCE/ECE cases with $b_e > b_i$). This suggests that a 'real brain' dynamics does not rely on a rigid mathematical cancellation of excitatory and inhibitory input strengths (i.e row-sum zero constraint) for its survival/ pursuits. Instead, it seemingly explores and exploits the specific interplay between connectivity range, synaptic strength asymmetry as well as the network's randomness to navigate the narrow channel of criticality, a mechanism that is far more sophisticated and regime-dependent than previously understood. Our results have potential relevance for achieving a specific brain function by altering  the complexity of the  network structure e.g. by external sources.

\acknowledgments

One of the authors (P.S.)  is grateful to Anusandhan national research foundation (ANRF), India for the financial support provided for the  research under Matrics grant scheme.




















\bibliographystyle{apsrev4-2}
\bibliography{reference}

\newpage

\newpage

\newpage

\clearpage

\newpage

\end{document}